\documentclass[sigplan, screen,table,usenames,dvipsnames]{acmart}

\renewcommand\footnotetextcopyrightpermission[1]{}

\AtBeginDocument{%
  }

\setcopyright{acmlicensed}
\copyrightyear{2027}
\acmYear{2027}
\acmDOI{XXXXXXX.XXXXXXX}

\usepackage{hyperref}
\usepackage{amsmath,amsfonts}
\usepackage{graphicx}
\usepackage{textcomp}
\usepackage{xcolor}
\usepackage{xspace}
\usepackage{multirow}
\usepackage[export]{adjustbox}
\usepackage{transparent}
\usepackage{enumitem}
\usepackage{hyphenat}
\usepackage{stfloats}
\usepackage{tikz}
\usetikzlibrary{shapes.geometric, arrows.meta, positioning, backgrounds, fit, calc, decorations.pathreplacing}

\usepackage{booktabs}
\usepackage{subcaption}

\usepackage{makecell}
\usepackage{graphicx}
\graphicspath{ {./images/} }
\usepackage{listings}
\usepackage{balance}
\usepackage{algorithm}
\usepackage{algorithmicx}
\usepackage[noend]{algpseudocode}
\usepackage{etoolbox}

\newcolumntype{?}{!{\color{blue}\vrule width 1pt}}

\definecolor{backcolour}{rgb}{0.95,0.95,0.92}
\definecolor{highlightgreen}{HTML}{80C904}
\definecolor{highlightred}{HTML}{BC544B}

\colorlet{lightyellow}{yellow!50}
\colorlet{lightapricot}{Apricot!50}
\colorlet{lightblue}{blue!30}
\colorlet{lightdandelion}{Dandelion!50}
\colorlet{lightthistle}{Thistle!50}

\newcommand{\metacomment}[3]%
{[{\color{#1}#2:\ #3}]}
\newcommand{\TODO}[1]{\metacomment{magenta}{TODO}{#1}}

\newcommand{\sys}{SABLE\xspace}

\newcommand{\eg}{{\em e.g.},\xspace}
 \newcommand{\ie}{{\em i.e.},\xspace}

\newcommand{\ddense}{$\delta$-dense\xspace}

\AtBeginEnvironment{algorithm}{\small}  % requires etoolbox (loaded in paper.tex)
\algnewcommand{\IIf}[1]{\State\algorithmicif\ #1\ \algorithmicthen}
\algnewcommand{\EndIIf}{\unskip\ \algorithmicend\ \algorithmicif}

\definecolor{codegreen}{rgb}{0,0.6,0}
\definecolor{codegray}{rgb}{0.5,0.5,0.5}
\definecolor{codepurple}{rgb}{0.58,0,0.82}

\lstdefinestyle{mystyle}{
    commentstyle=\color{codegreen},
    keywordstyle=\color{magenta},
    numberstyle=\tiny\color{codegray},
    stringstyle=\color{codepurple},
    basicstyle=\ttfamily\scriptsize,
    breakatwhitespace=false,
    breaklines=true,
    captionpos=b,
    keepspaces=true,
    numbers=left,
    numbersep=5pt,
    showspaces=false,
    showstringspaces=false,
    showtabs=false,
    tabsize=2
}

\setlist{leftmargin=*}

\newwrite\appendixstartfile

\begin{document}

\title{Bring Your Own Formats and Kernels: Composable Abstractions for Sparse Matrix Computation}

\begin{abstract}
  Real-world sparse matrices often feature multiple forms of structured sparsity---rectangular dense blocks, diagonal bands, and scattered entries---that no single storage format can efficiently exploit.
Hybrid formats address this by storing each subregion of a matrix in its most efficient form. Existing hybrid approaches, however, only support fixed sets of formats and kernels, so incorporating a new representation or kernel requires modifying their internals.
We present \sys{}, a framework that lets users build bespoke hybrid formats compositionally through a \emph{plan--extract--dispatch} interface.
Users define \emph{extractors} that carve a matrix into format-specific regions and \emph{kernels} that emit specialized C code for each region; \sys{} assembles these pieces into a single program specialized to the target matrix at compile time.
Both components are independent and composable, so a new format automatically integrates with all existing kernels without any changes to the framework.
We demonstrate this extensibility by introducing VDIA, a novel format for diagonal bands of non-uniform length, and composing it to build two new hybrid formats---VDIA+CSR and VDIA+VBR+CSR.
We evaluate \sys{} on SpMV and SpMM using matrices from the SuiteSparse benchmarks, demonstrating geometric-mean speedups over the best fully-sparse baselines of $1.10\times/1.20\times$ (SpMV/SpMM) for VBR+CSR, and $1.14\times/1.31\times$ for VDIA+CSR, with the full VDIA+VBR+CSR composition yielding a further $1.08\times/1.25\times$ over VBR+CSR.

\end{abstract}


\author{Pratyush Das}
\email{das160@purdue.edu}
\affiliation{%
	\institution{Purdue University}
	\city{West Lafayette}
	\state{Indiana}
	\country{USA}
}

\author{Amirhossein Basareh}
\email{abasareh@purdue.edu}
\affiliation{%
	\institution{Purdue University}
	\city{West Lafayette}
	\state{Indiana}
	\country{USA}
}

\author{Artem Pelenitsyn}
\email{apelenit@purdue.edu}
\affiliation{
	\institution{Purdue University}
	\city{West Lafayette}
	\state{Indiana}
	\country{USA}
}

\author{Kirshanthan Sundararajah}
\email{kirshanthans@vt.edu}
\affiliation{
	\institution{Virginia Tech}
	\city{Blacksburg}
	\state{Virginia}
	\country{USA}
}

\author{Milind Kulkarni}
\email{milind@purdue.edu}
\affiliation{
	\institution{Purdue University}
	\city{West Lafayette}
	\state{Indiana}
	\country{USA}
}

\author{Ben Delaware}
\email{bendy@purdue.edu}
\affiliation{%
	\institution{Purdue University}
	\city{West Lafayette}
	\state{Indiana}
	\country{USA}
}

\maketitle

\section{Introduction}
\label{sec:introduction}

% Sparse matrices are ubiquitous
\textit{Sparse matrices} --- matrices comprised of mostly zeroes --- commonly appear in numerous domains, including machine learning, graph processing, and scientific computing applications~\cite{sellp, graphcentrality, gnnperf, gmres}.
% Different formats exploit different structures.
Specialized representations that exploit the limited number of non-zero elements in a sparse matrix provide immediate storage and computational benefits over standard, dense representations.
Using indirection arrays to store non-zero elements~\cite{csryale, coo}, for example, can shrink the memory needed to store a sparse array by an order of magnitude, while specialized kernels can greatly accelerate zero-preserving operations, e.g., matrix-vector and matrix-matrix multiplication, over this representation.
In addition, many matrices feature some form of \textit{structured sparsity}, where non-zero elements appear in regular patterns instead of being randomly distributed.
Decades of research have produced a rich landscape of representations, each of which targets a particular form of structure, including fixed-size blocks~\cite{bcsr}, diagonal bands~\cite{saad2003iterative}, and row-length uniformity~\cite{ellpack}.
% There are many different forms of structured sparsity which representations can take advantage of.

\begin{figure}[t!]
    \centering
    \begin{subfigure}{0.48\columnwidth}
        \centering
        \includegraphics[width=\textwidth]{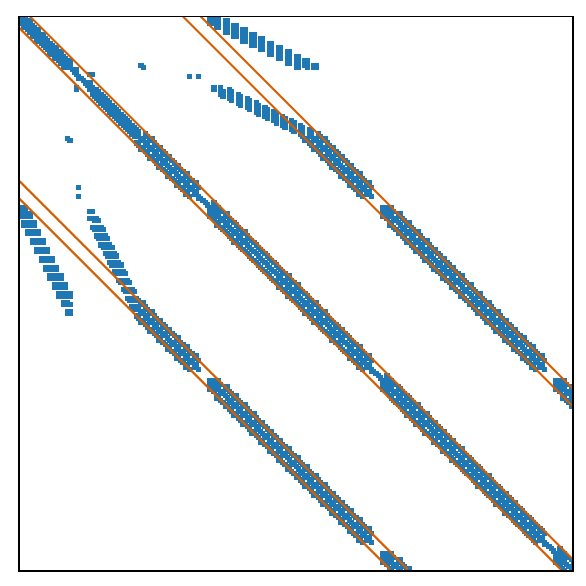}
        \caption{\textit{ex18}~\cite{SuiteSparse}}
        \label{fig:motivating_matrix_vdia}
      \end{subfigure}
      \hfill
    \begin{subfigure}{0.48\columnwidth}
        \centering
        \includegraphics[width=\textwidth]{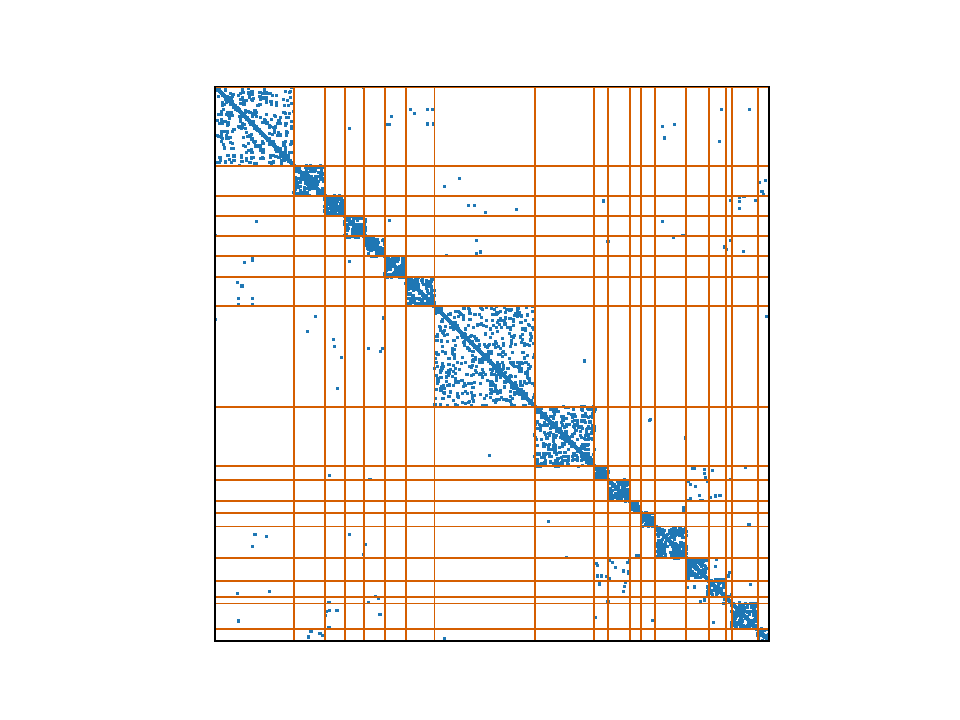}
        \caption{\textit{bcspwr06}~\cite{SuiteSparse}}
        \label{fig:motivating_matrix_vbr}
    \end{subfigure}
    \caption{Two SuiteSparse~\cite{SuiteSparse} matrices that feature different forms of structured sparsity; blue dots are non-zero elements and dense subregions are demarcated in orange.}
      % (\subref{fig:motivating_matrix_vbr}) \textit{bcspwr06} contains dense rectangular blocks (demarcated in orange) that VBR captures, with the remaining scattered entries left to CSR.
      % (\subref{fig:motivating_matrix_vdia}) \textit{ex18} (shown zoomed in to its top-left corner) contains several diagonals at distinct offsets (demarcated in orange) that VDIA captures, with the remaining scattered entries left to CSR.}
    \label{fig:motivating_matrix}
    \vspace{-10pt}
\end{figure}

As a simple example, consider the two matrices shown in \autoref{fig:motivating_matrix}, each of which exhibits a different form of structured sparsity.
The matrix on the left (\texttt{ex18}) contains several diagonal bands of (mostly) non-zero elements of varying lengths at distinct offsets, while the matrix on the right (\texttt{bcspwr06}) includes several rectangular blocks of varying sizes that consist of (mostly) non-zeroes.
The DIA~\cite{saad2003iterative} format targets subregions similar to those in \texttt{ex18}, while the Variable Block Row (VBR)~\cite{SPARSKIT} format is designed to capture subregions of the latter form.
Both of these representations eliminate the memory overhead imposed by indirection arrays, and support specialized kernels that can exploit the structured memory access patterns and vectorization opportunities provided by the regular shape of these regions.
Crucially, these regions do not need to be fully dense, e.g., the center block in \autoref{fig:motivating_matrix_vbr}, or perfectly align with a particular shape, e.g., the lower-left band in \autoref{fig:motivating_matrix_vdia}, to enjoy these benefits, as long as they outweigh the cost of storing and computing over a few zero elements.

Notably, a \textit{single} matrix can feature different forms of structured sparsity -- both matrices in \autoref{fig:motivating_matrix}, for example, include some non-zero elements outside their structured regions that are more effectively represented by a standard sparse format.
\textit{Hybrid formats} attempt to address this by decomposing a matrix into subregions and storing each in its most efficient representation; having done so, it is straightforward to dispatch computations over each region to an appropriate kernel~\cite{lav, tilespmv, stile, uzp}.
Unfortunately, existing hybrid approaches only support fixed sets of representations: LAV~\cite{lav}, for example, can only decompose matrices into dense and CSR regions, TileSpMV~\cite{tilespmv} draws from among a fixed menu of formats, and STile~\cite{stile} requires modifying the compiler to add new formats.
As a consequence, incorporating new representations or kernels into these existing frameworks --- say, a variable-length diagonal format, or a vendor BLAS kernel --- requires modifying their internals, leaving a gap between the space of useful combinations of formats and what these systems can express: to the best of our knowledge, none of these can efficiently capture the sparsity structure shown in \autoref{fig:motivating_matrix_vdia}, for example.

To address this gap, this paper presents a new framework, called \sys{}, that allows users to build bespoke hybrid formats in a \textit{composable} fashion, by combining different representations using a simple \emph{plan--extract--dispatch} workflow.
Users supply \sys{} with a target matrix and operation, as well as a high-level \textit{plan} identifying the sorts of structures the user thinks may appear in that matrix.
For each specified structure, \sys{} automatically \textit{extracts} subregions of the target matrix exhibiting that structure into an appropriate representation.
Finally, \sys{} generates a program that computes the target operation by \textit{dispatching} computations over each subregion to a kernel specialized to its structure.
This composability comes at a price: in its current form, \sys{} supports only operations that accumulate their results through an associative and commutative operator, e.g., the additions underlying matrix-vector and matrix-matrix multiplication, ensuring that each subregion can be computed independently and the partial results can be accumulated into the shared output in any order.
This workflow naturally aligns with the workloads found in many domains, e.g., pruned neural networks~\cite{venom}, graph neural networks~\cite{gnnperf}, and iterative solvers~\cite{gmres}, where the non-zero values of a sparse matrix change across invocations but its sparsity patterns remain fixed.
For such matrices, \sys{} can inspect a matrix's structure once and then generate a specialized program in which all dispatch decisions, loop bounds, and array offsets are resolved at compile time.

%\TODO{Are we the first? Is arbitrary actually true? Does \sys have restrictions? (yes, semiring? commutative and associative ops?) Are there constraints on what constitutes a valid format?}
Moreover, this workflow can easily be extended to support new forms of structured sparsity.
As a simple example, \sys{} includes a novel form of the structure shown in \autoref{fig:motivating_matrix_vdia}, whose underlying representation, Variable DIAgonal (VDIA), generalizes DIA similarly to how VBR generalizes fixed-size blocks.
Unlike fixed-band DIA, which must pad every band to the full length of the matrix, VDIA stores each band at its actual extent.
Integrating this structure into \sys{} was a simple matter of providing an implementation of an interface whose methods explain how to carve out matrix subregions with that structure and how to generate specialized C code for its representation.
% Benefits of composition.
Importantly, each new component naturally composes with all existing kernels (and vice versa) without any changes to \sys{}.
This design enables novel hybrid format combinations --- VDIA+CSR and VDIA+VBR+CSR --- that arise naturally from composing independently defined extractors and kernels.

In summary, this paper makes the following contributions:%
\begin{itemize}[leftmargin=*]
    \item \textbf{A compositional framework for building hybrid formats.}
      We present the first system that allows users to define and compose sparse storage formats with kernels for a specific matrix, through a \emph{plan--extract--dispatch} interface.
      Unlike prior hybrid approaches that hard-code specific format/kernel combinations, \sys{} treats both formats and kernels as user-extensible components.
      %(\autoref{sec:framework})
    \item \textbf{A codegen-based approach to composing hybrid formats.}
      \sys{} represents formats and kernels as independent, self-contained components, generating code based on a user-defined plan and specialized to a target matrix, rather than threading every combination through a single generic program.
      We present a two-stage compilation pipeline that resolves all dispatch decisions, loop bounds, and array offsets at compile time.
      %In particular, staging turns the structural metadata of variable-sized formats such as VBR and VDIA into compile-time constants, so a kernel author can express them as cleanly as fixed-size formats while retaining their ability to adapt to irregular structure.
      %(\autoref{sec:staging})
    %\item \textbf{A new variable-length diagonal format.} Unlike fixed-band DIA, which must pad every band to the full length of the matrix, VDIA stores each band at its actual extent. \TODO{Does this have to be a contribution?}
    \item \textbf{Novel hybrid formats via composition.}
      To demonstrate that \sys{} accommodates new representations, we introduce VDIA, a novel format for diagonal bands of non-uniform length, which we use to define novel compositional hybrid formats, VDIA+CSR and VDIA+VBR+CSR, without any changes to the underlying framework. These formats have not previously appeared in the literature, showcasing that useful formats arise naturally from composing independently defined extractors and kernels.
      %(\autoref{sec:implementation})
\end{itemize}

We evaluate \sys{} on SpMV and SpMM over matrices from the SuiteSparse benchmark suite~\cite{SuiteSparse}, demonstrating geometric-mean speedups of $1.10--1.31\times$ over the best fully-sparse baselines (\autoref{sec:evaluation}), using \sys{}'s built-in extractors to automatically detect dense blocks and diagonal bands, enabling an end-to-end pipeline from matrix to specialized executable.

\section{Overview}%
\label{sec:overview}

\begin{figure}[t]
  \centering
  \scalebox{.78}{%
  \begin{tikzpicture}[
    phaseRegion/.style={
        draw,
        dashed,
        color=gray!50,
        fill=gray!2,
        rectangle,
        rounded corners=2pt,
        inner sep=10pt
    },
    block/.style={
        draw,
        rectangle,
        minimum width=2.7cm,
        minimum height=0.5cm,
        align=center,
        fill=green!8,
        line width=1.1pt,
        rounded corners=3pt
    },
    smallblock/.style={
        draw,
        rectangle,
        minimum width=0.9cm,
        minimum height=0.75cm,
        align=center,
        fill=green!8,
        line width=1.1pt,
        rounded corners=3pt
    },
    inputblock/.style={
        draw,
        rectangle,
        minimum width=1.5cm,
        minimum height=0.5cm,
        align=center,
        fill=orange!10,
        line width=1.1pt,
        rounded corners=3pt
    },
    smallinput/.style={
        draw,
        rectangle,
        minimum width=0.7cm,
        minimum height=0.75cm,
        align=center,
        fill=orange!10,
        line width=1.1pt,
        rounded corners=3pt
    },
    fileblock/.style={
        draw,
        color=black,
        fill=blue!8,
        text=black,
        rectangle,
        minimum width=1.8cm,
        minimum height=0.5cm,
        align=center,
        line width=1.2pt,
        rounded corners=2pt
    },
    arrow/.style={
        ->,
        >={Stealth[scale=1]},
        line width=1pt
    },
    font=\sffamily\small
]

    % =========================================================================
    % 1. INPUT + UNROLLED FORMAT EXTRACTION ROW
    % =========================================================================
    % Middle X is the anchor; other nodes hang off it left/right.
    \node (x2) [smallblock] {$X_2$};
    \node (r1) [smallinput, left=0.35cm of x2] {$R_1$};
    \node (x1) [smallblock, left=0.35cm of r1] {$X_1$};
    \node (input_mtx) [inputblock, left=0.6cm of x1] {\textbf{Sparse}\\\textbf{Matrix}};
    \node (r2) [smallinput, right=0.35cm of x2] {$R_2$};
    \node (dots) [right=0.25cm of r2, inner sep=2pt] {$\cdots$};
    \node (x3) [smallblock, right=0.25cm of dots] {$X_n$};
    \node (zero) [smallinput, right=0.35cm of x3] {$O$};

    % (Format Extraction sub-region is drawn below in the background-layer scope,
    %  so its dashed border sits on top of the surrounding PLANNING region.)

    % =========================================================================
    % 2. PLANNING PHASE COMPONENTS
    % =========================================================================
    \node (kernels) [block, below=1.1cm of x2] {
        \textbf{Dispatching}\\
        {\small $\{\text{format} \mapsto \text{kernel}\}$}
    };

    \node (compile_call) [block, below=0.7cm of kernels] {
        \textbf{Code Generation}
    };

    % =========================================================================
    % 3. ARTIFACT BRIDGE
    % =========================================================================
    \node (sabledata) [fileblock, right=0.8cm of compile_call] {
        \texttt{.sabledata}
    };
    \node (c_code) [fileblock, left=0.8cm of compile_call] {
        \texttt{.c}
    };

    % =========================================================================
    % 4. EXECUTION PHASE COMPONENTS
    % =========================================================================
    \node (executor_build) [block, below=1.2cm of compile_call] {
        \textbf{Compile \& Link}
    };

    \node (deps) [inputblock, left=0.5cm of executor_build] {
        \textbf{Libraries}\\
        MKL, BLAS
    };

    \node (executor_run) [block, below=0.7cm of executor_build] {
        \textbf{Binary Execution}
    };

    \node (output_data) [inputblock, below=0.7cm of executor_run] {
        \textbf{Output Matrix}
    };

    \node (user_rhs) [inputblock, left=.5cm of executor_run] {\textbf{Dense Vector}};

    % =========================================================================
    % 5. DIAGRAM LEGEND (AT THE VERY BOTTOM)
    % =========================================================================
    \node (legend_box) [draw, color=gray!60, fill=white, rounded corners=4pt,
                         below=0.2cm of output_data,
                         minimum width=9.4cm, minimum height=1.5cm] {};
    \node (legend_label) [right=0.15cm of legend_box.north west, yshift=-5mm, font=\sffamily\bfseries\small, color=gray!100] {LEGEND};
    \node (leg1) [inputblock, minimum height=0.4cm,
                  right=1.8cm of legend_label.west, font=\sffamily\small] {Inputs / Outputs};
    \node (leg2) [block, minimum width=1cm, minimum height=0.4cm,
                  right=0.3cm of leg1, font=\sffamily\small] {Operations};
    \node (leg3) [fileblock, minimum width=1cm, minimum height=0.4cm,
                  right=0.3cm of leg2, font=\sffamily\small] {Files};
    \node (leg5) [smallinput, minimum width=0.55cm, minimum height=0.4cm,
                  below=1mm of legend_label, font=\sffamily\small, xshift=-4mm]
                  {$R_i$};
    \node (leg5txt) [right=0.1cm of leg5, font=\small] {--- residual matrix};
    \node (leg4) [smallblock, minimum width=0.55cm, minimum height=0.4cm,
                  right=0.4cm of leg5txt, font=\sffamily\small] 
                  {$X_i$};
    \node (leg4txt) [right=0.1cm of leg4, font=\small] {--- extractor};
    \node (leg6) [smallinput, minimum width=0.55cm, minimum height=0.4cm,
                  right=0.4cm of leg4txt, font=\sffamily\small] 
                  {$O$};
    \node (leg6txt) [right=0.1cm of leg6, font=\small] {--- zero matrix};

    % =========================================================================
    % VISUAL PHASE REGIONS (BACKGROUND + INTERNAL HEADINGS)
    % =========================================================================
    \begin{scope}[on background layer]
        % Invisible sizing helper: defines the FE area's bbox so planning_region can fit it.
        \node (fe_box) [rectangle, fit=(x1)(zero), minimum height=1.3cm, yshift=2mm] {};

        % Outer Planning region — drawn FIRST so its fill sits behind the inner FE region.
        \node (planning_region) [phaseRegion, fit=(fe_box)(kernels)(compile_call), yshift=2mm] {};
        \node [font=\sffamily\bfseries\small\color{gray!100},
               anchor=north west, xshift=4pt, yshift=0pt]
              at (planning_region.north west) {PLANNING};

        % Inner Format Extraction sub-region around just the X-R sequence
        \node (fe_region) [phaseRegion, inner sep=.5mm, fit=(fe_box)] {};
        \node [font=\sffamily\bfseries\small\color{gray!100},
               anchor=north west, xshift=4pt]
              at (fe_region.north west) {Format Extraction};

        % Execution Phase Region
        \node (execution_region) [phaseRegion, fit=(user_rhs)(deps)(executor_build)(executor_run), minimum height=3cm] {};
        \node [font=\sffamily\bfseries\small\color{gray!100},
               anchor=north west, xshift=5pt]%, yshift=-12pt]
              at (execution_region.north west) {EXECUTION};
    \end{scope}

    % =========================================================================
    % LOGICAL PIPELINE FLOW
    % =========================================================================
    % Horizontal extraction sequence: matrix -> X1 -> R1 -> X2 -> R2 -> ... -> X3 -> O
    \draw [arrow] (input_mtx) -- (x1);
    \draw [arrow] (x1) -- (r1);
    \draw [arrow] (r1) -- (x2);
    \draw [arrow] (x2) -- (r2);
    \draw [arrow] (r2) -- (dots);
    \draw [arrow] (dots) -- (x3);
    \draw [arrow] (x3) -- (zero);

    % Each extractor stage feeds Kernel Binding with its own format
    \draw [arrow] (x1.south) -- node[midway, right, font=\sffamily\small]
        {format$_1$} (kernels.north west);
    \draw [arrow] (x2.south) -- node[midway, right, font=\sffamily\small]
        {format$_2$} (kernels.north);
    \draw [arrow] (x3.south) -- node[midway, right, font=\sffamily\small, xshift=1mm]
        {format$_n$} (kernels.north east);

    \draw [arrow] (user_rhs) -- (executor_run.west);

    \draw [arrow] (kernels) -- (compile_call);

    % Lines to artifacts
    \draw [arrow, <-] (sabledata) -- (compile_call);
    \draw [arrow, <-] (c_code) -- (compile_call);

    % Downward phase hand-off
    \draw [arrow] (compile_call) -- (executor_build);
    \draw [arrow] (deps) -- (executor_build);

    % Vertical side lines connecting artifacts down to Build
    \draw [arrow] (sabledata) |- (executor_run);
    \draw [arrow] (c_code) -- (executor_build.north west);

    \draw [arrow] (executor_build) -- (executor_run);
    \draw [arrow] (executor_run) -- (output_data);

\end{tikzpicture}
  }
  \vspace{-2pt}
  \caption{Overview of the \sys{} framework.}%
  \label{fig:sable-overview}
  \vspace{-10pt}
\end{figure}

%\TODO{Update overview diagram}

Every sparse matrix format makes an (implicit) assumption about the \textit{structure} of a matrix, i.e., where its non-zeros live.
This assumption then informs the choice of the best in-memory \emph{representation} of that structure.
DIA, for example, assumes non-zeros cluster along a few diagonals and stores one dense array per diagonal; BCSR assumes fixed-size dense blocks stored column-major; CSR assumes nothing and uses compressed indirection arrays.
\sys{} is organized around these two separate, but connected concerns: it first identifies subregions of a sparse matrix exhibiting user-specified  structure, e.g., rectangular dense blocks, diagonal bands, and then stores each piece using the representation best suited for that structure.
This representation also determines how to \emph{compute} over each piece: the loop structure, the memory access patterns, and the library routines it supports.

\autoref{fig:sable-overview} provides a high-level overview of \sys{}'s workflow.
\sys{} first finds the structured subregions of a sparse matrix using \emph{extractors}, functions that identify the subregions of a matrix exhibiting a particular form of structure and store them in a representation suited to that structure.
Each extractor produces two results: a \emph{format object} that stores the non-zero values of the extracted subregions in the target format;  and a \emph{residual matrix} holding any non-zero values outside those regions.
This allows \sys{} to implement the search for structured subregions as a sequence of extractors, where each claims the non-zeros that fit its target structure and leaves the rest to its successors, carving a heterogeneous matrix into a composition of homogeneous pieces.
This strategy is essential to \sys{}'s composability story, as each extractor operates independently of what came before and what follows.
Format objects are similarly agnostic to how they were constructed.
A CSR region might be produced by a generic ``claim everything that is left'' extractor or by a specialized extractor that sweeps non-zeros off a matrix's lower diagonals; either way the resulting CSR object is identical, and can be processed by any CSR kernel.

%\TODO{Refer to overview diagram after it is updated}
\iffalse
\paragraph{Formats as the boundary}
Formats are the interface in the middle of the pipeline, and they are just data.
Once a region has been extracted, its format object is all that remains of it: downstream stages neither know nor care which extractor produced the format, what structure it was carved from, or what the rest of the matrix looked like.
For instance, a CSR region might be produced by a generic ``claim everything that is left'' extractor or by a specialized extractor that sweeps non-zeros off a matrix's lower diagonals; either way the resulting CSR object is identical, and any CSR kernel can consume it.
Conversely, the same structure can be stored in more than one representation, and the choice can be revisited without touching the code that found the structure.

\paragraph{Dispatch: operations over formats}
\fi

Each format object is equipped with a \emph{kernel} which emits specialized C code that computes a target operation over that representation.
A kernel may emit loop nests specialized to the subregion's metadata, make calls into open-source or vendor libraries such as OpenBLAS or MKL, or a mixture of both, choosing per region while the code is being generated.
\sys{} assembles the generated pieces into a single C program that computes the target operation over the full matrix, with each piece accumulating its partial result into a shared output.
%\TODO{Refer to overview diagram after it is updated}
\sys{} thus follows the inspector--executor paradigm~\cite{inspector93}: the extractors form the inspector, which analyzes the matrix structure once, while the generated program serves as the executor, which can be reused across computations that share that structure.
Extractors and kernels only interact via format objects.
A new extractor is compatible with any kernel for its target representation, and a new kernel is similarly compatible with every extractor producing its target representation, without any changes to the framework.
Thus, \sys{} imposes limited constraints on extractors and kernels: each kernel must accept the extracted subregion's type, all kernels must agree on the top-level matrix operation, and that operation must decompose over subregions.

We pause here to briefly highlight the distinguishing features of \sys{}'s approach; \autoref{sec:related_work} provides a fuller comparison with prior work.
To the best of our knowledge, no other system allows this level of composability for hybrid sparse computation.
Existing hybrid approaches~\cite{lav, tilespmv, stile, uzp} bake in specific format/kernel pairs, so adding a new format or dispatch strategy requires modifying their internals.
Compiler abstractions such as TACO~\cite{taco}, SparseTIR~\cite{sparsetir}, and UniSparse~\cite{unisparse} are expressive languages for \emph{representations}---how coordinates and values are encoded---but discovering the best representation of a particular matrix's subregions falls outside their scope.
Inspector--executor systems~\cite{inspector93, registertiling, kazemcodelet} do inspect a concrete matrix, but each couples its inspection to one hard-coded specialization strategy.

\iffalse
\paragraph{Why the separation matters}
Separating structure from representation is what makes \sys{} compositional.
Extractors and kernels never interact directly; they meet only at format objects.
A new extractor therefore composes with every existing kernel for the representation it targets, and a new kernel composes with every extractor that produces the format it accepts, without any changes to the framework.
The framework imposes no constraints beyond type compatibility: a kernel must accept the format's type, and all kernels in a plan must agree on the operation.
To the best of our knowledge, no prior system allows this level of composability for hybrid sparse computation.
Existing hybrid approaches~\cite{lav, tilespmv, stile, uzp} bake in specific format/kernel pairs, so adding a new format or dispatch strategy requires modifying the system internals.
\TODO{Cite Mosaic}

\paragraph{Discussion}
Compiler abstractions such as TACO~\cite{taco}, SparseTIR~\cite{sparsetir}, and UniSparse~\cite{unisparse} are expressive languages for \emph{representations}---how coordinates and values are encoded---but discovering which regions of a particular matrix instance warrant which representation falls outside their scope.
Inspector--executor systems~\cite{inspector93, registertiling, kazemcodelet} do inspect a concrete matrix, but each couples its inspection to one hard-coded specialization strategy.
\fi

\section{Core Abstractions}%
\label{sec:plan_interface}

Users drive \sys{} through a single \lstinline{plan} object, which scripts the pipeline of \autoref{sec:overview}: the plan decides how to break a matrix up, applies extractors to a concrete matrix to carve it into formats, and dispatches each format to a kernel.
\autoref{fig:plan} shows an example plan for an SpMM operation which extracts dense blocks into the VBR format and stores the scattered remainder in CSR.
Every plan consists of an initialization, extraction, and dispatch phase:

\begin{figure}
\centering
\begin{minipage}{0.85\linewidth}
\begin{lstlisting}[language=Python,xleftmargin=\parindent,escapechar=&]
matrix = Matrix("matrix.mtx")
plan = Plan(matrix, artifact_dir="./artifacts")
plan.rhs(DenseInput.matrix("rhs.matrix",
    shape=(matrix.ncols, 512),
    layout=DenseLayout.ROW_MAJOR)) &\vspace{3pt}&
# Carve the matrix into formats.
vbr = plan.extract(BlockExtractor())
csr = plan.extract(CSRExtractor()) &\vspace{3pt}&
# Bind a kernel to each format.
plan.dispatch(vbr, MixedVBRSpmm())
plan.dispatch(csr, NaiveCSRSpmm()) &\vspace{3pt}&
# Generate, build, and run the C program.
executor = plan.compile(filename="my_spmm")
executor.build()
output = executor.run()
\end{lstlisting}
\end{minipage}
\vspace{-5pt}
\caption{Example of a user plan for SpMM in \sys{}}%
\label{fig:plan}
\vspace{-5pt}
\end{figure}

\paragraph{Initialize}
The user creates a plan from a sparse matrix and declares the dense operand (\lstinline{plan.rhs})---its location on disk, shape, and layout.
Internally, the plan maintains a \emph{residual matrix}---initially the full input matrix---and an ordered list of dispatches.
The residual tracks the non-zeros that have not yet been claimed by any extractor.

\paragraph{Extract}
Each call to \lstinline{plan.extract(extractor)} hands the current residual to the extractor, which returns a format object containing the region it claimed and a new residual with those non-zeros removed.
Successive extractions shrink the residual, carving structured regions out of the matrix one extractor at a time.
The plan type-checks each step: the returned format must match the extractor's declared \lstinline{produces} type, and the residual may never gain non-zeros.

\paragraph{Dispatch}
Each call to \lstinline{plan.dispatch(format, kernel)} binds a kernel to an extracted format.
The plan verifies that the kernel's \lstinline{accepts} type matches the format, and that all dispatched kernels agree on the operation (SpMV or SpMM in the current implementation; users can add their own operations).
Before generating code, the plan checks completeness: every non-zero of the input matrix must have been claimed by some extraction, and every extracted format must have a kernel.
Because extractors only claim non-zeros that fit their target structure, a plan ends with a catch-all extractor whose residual array is always empty, e.g., \lstinline{CSRExtractor}, whose format can capture arbitrary non-zeros. The user can dispatch structured representations of subregions to any compatible kernel.
For example, the result of a VBR extractor can be dispatched to a generated loop nest, an MKL BLAS kernel, or a mixed kernel that chooses per block; while CSR can be dispatched to SpV8, MKL, UZP, or a generated loop.

Because extractors and kernels are ordinary Python objects, users can define new ones without modifying others.
We now describe how each component is instantiated.

\subsection{Formats}
\label{sec:formats}

A \emph{format} describes how an extracted region is stored.
Formats are implemented as a Python dataclass that holds the number of columns and rows, and optional, format-specific fields, e.g., for structural metadata.
\sys{} ships with three built-in formats spanning a spectrum of useful structures: \textbf{CSR}, which makes no assumptions about structure and can represent any sparse matrix; \textbf{VBR}, which stores variable-sized dense blocks; and \textbf{VDIA}, which stores diagonal bands.

A format's fields divide into two classes that differ in \emph{when} they are consumed.
Plain Python values---scalars and lists---exist only while code is being generated: kernels read them to make decisions and splice them into the emitted source as constants.
Fields wrapped in \lstinline{Rep}, by contrast, name arrays that the generated program owns at run time: \sys{} assigns each \lstinline{Rep} a unique C identifier, infers its C type, serializes its contents to a data file, and emits code that loads the array at startup.
The format author thereby chooses, field by field, what the generated program sees.
This is how \sys{} expresses variable-sized formats as cleanly as fixed-size ones: the indirection that VBR and VDIA would otherwise pay for at run time is resolved while the code is generated.
Users define new formats by subclassing \lstinline{Format}, adding fields appropriate for the target storage scheme.

\subsection{Extractors}%
\label{sec:extractors}

An \emph{extractor} is a rule for finding regions of a particular structure in a matrix and packing them into a format.
Each extractor declares the format type it produces, enabling the plan to enforce type compatibility at each step.
An extractor is expected to be greedy: it claims as much of its target structure as it can find, leaving behind only the non-zeros that do not fit that structure.
The simplest case is a CSR extractor that claims all remaining non-zeros.
More complex extractors call external tools.
For instance, our \lstinline{BlockExtractor} invokes a C++-based partitioner (\autoref{sec:block_partitioner}) to find variable-sized \ddense blocks and returns a VBR format;
our \lstinline{BandExtractor} extracts diagonal regions (\autoref{sec:band_partitioner}) and returns a VDIA format.
In both cases, the extractor also returns a new residual matrix with the extracted non-zeros removed.
%A user can integrate a new storage format into \sys{} by implementing a new extractor, without modifying any other component of the system.
A simple example of instantiating an extractor is given in Appendix~\ref{sec:inst-extractor}.

\subsection{Kernels}%
\label{sec:kernels}

A \emph{kernel} maps a format object to a C code fragment for the target operation.
The format's fields resolve at code-generation time according to their class (\autoref{sec:formats}): plain values evaluate to Python integers or lists and are inlined as constants in the emitted C source, while \lstinline{Rep}-wrapped fields evaluate to C identifiers that name the corresponding run-time arrays.
A kernel, therefore, writes code in which structural parameters are already resolved to constants---loop bounds are literal integers, array names are fixed---with no run-time dispatch or indirection.
A simple example of instantiating a kernel is given in Appendix~\ref{sec:inst-kernel}.

More sophisticated kernels exploit the metadata available while code is being generated.
Our \lstinline{MixedVBRSpmm} iterates over a VBR format's blocks and emits, for each block, either a call to \lstinline{cblas_dgemm} or a loop nest with constant bounds, depending on the block's dimensions: blocks whose smallest dimension is at least 8 and whose aspect ratio is at most 100 are sent to BLAS, while smaller or thinner blocks---for which the BLAS call overhead outweighs its benefit---fall back to the generated loop.
The decision is made once per block, while the code is generated, and baked into the program.

By default, every snippet a kernel emits is inlined into the generated program's \lstinline{main} function.
For formats with many regions this can bloat the source: a VBR matrix with hundreds of blocks would repeat the same loop nest hundreds of times.
A kernel can instead wrap a snippet in \lstinline{out_of_line(...)}, which hoists it into a parameterized helper function.
\sys{} automatically adds the output and input vectors and any referenced \lstinline{Rep} arrays to the helper's signature; scalar values that vary between calls---bounds, offsets, and bandwidths---are passed as explicit parameters.
Snippets that share a name and body share one helper, so a kernel that emits the same loop for every block produces one function and many calls.
Plain strings stay inline, so a kernel can mix inline library calls with out-of-line loop bodies, as \lstinline{MixedVBRSpmm} does.

Kernels that wrap external libraries can declare headers, compiler flags, linker flags, and environment variables the generated binary needs.
For example, a user can dispatch CSR to a new vectorized implementation or VBR to a custom SIMD routine by writing a new kernel class.
%Like extractors, kernels are user-extensible: to dispatch CSR to a new vectorized implementation, or VBR to a custom SIMD routine, the user writes a new kernel class without modifying other components.

\subsection{From a Plan to an Executable}%
\label{sec:executable}

Compilation (\lstinline|plan.compile()| in \autoref{fig:plan}) turns a complete plan into a C program in a single pass over its dispatches.
The compiler first checks that the residual is empty and every extracted format has a kernel, then binds every \lstinline{Rep} across all formats to a unique C identifier and serializes its values into a binary data file (\lstinline{.sabledata}).
The compiler then invokes each kernel's \lstinline{emit_call} operation with its concrete format object and assembles the emitted snippets, in dispatch order, into one C source file whose \lstinline{main} loads the \lstinline{.sabledata} arrays and the dense operand, executes each dispatch, and accumulates the partial results into a shared output buffer.
The last two steps are deliberately separate.
Building (\lstinline{executor.build()}) invokes the C compiler with the flags collected from the dispatched kernels, and running (\lstinline{executor.run()}) executes the resulting binary.
This separation lets users independently inspect, modify, or recompile the generated code.

Two key properties --- no run-time format dispatch, and dependence only on sparsity structure rather than values --- follow directly from resolving structure at code-generation time (\autoref{sec:formats}) and storing values in the separate \lstinline{.sabledata} file. The same executable can therefore be reused across matrices sharing a sparsity pattern, e.g., as commonly happens in iterative solvers and graph neural networks.

\section{Implementation}%
\label{sec:implementation}

Our implementation of \sys{} includes built-in support for the two forms of structured sparsity shown in \autoref{sec:introduction}, i.e., rectangular dense blocks and diagonal band regions. %
In both cases, \sys{} supports extracting variable-sized subregions: block and segment boundaries adapt to the matrix's actual structure rather than imposing a fixed tile size. %
Per \autoref{sec:plan_interface}, users can freely extend the framework to support additional kinds of structured sparsity.

\subsection{Block Extractor}
\label{sec:block_partitioner}

\lstinline{BlockExtractor} is an extractor that identifies variable-sized mostly dense blocks in a sparse matrix that can be efficiently represented using SPARSKIT's VBR format~\cite{SPARSKIT}. %
It delegates the search for subregions with this structure to a partitioner written in C++, parses the resulting block coordinates, and represents the blocks using the VBR format.
VBR lets block boundaries adapt to the matrix's actual nonzero structure, covering dense regions with minimal fill-in.\footnote{\autoref{sec:vbr_format} in the appendix describes VBR's in-memory representation.}

The extraction algorithm is a heuristic search for non-overlapping high-density rectangular blocks.
Finding the optimal set of such blocks is NP-hard~\cite{npcomplete}, so the heuristic uses three design elements:
\begin{itemize}[leftmargin=*]
\item[$-$] A minimum density threshold~$\delta$ (default 50\%): a block is claimed only if it is \ddense{}, where $\delta$ reflects the break-even point where the cost of computing over zero elements is offset by the performance of dense kernels.
\item[$-$] A minimum area threshold $A_{\min}$ (default 2500): this ensures blocks are large enough to amortize the overhead of switching between dense and sparse code paths. The block search requires the outer dimension to be at least $\lceil\sqrt{A_{\min}}\,\rceil$; any span shorter than this cannot reach the area threshold, so it is skipped.
\item[$-$] A scoring function, $\mathrm{score}(B)=\mathrm{area}(B)\cdot(\mathrm{density}(B)-0.5)^{1.5}$, that prefers denser blocks while still accounting for size; the super-linear exponent favors blocks with substantially higher density over larger, less dense blocks.
\end{itemize}

%% Custom algorithmic block for parallel for
\algdef{SE}{ParFor}{EndParFor}[1]{\textbf{parallel for} #1 \textbf{do}}{\textbf{end parallel for}}

\begin{algorithm}[t]
\caption{Partitioner}\label{alg:partitioner}
\begin{algorithmic}[1]
\Require CSR and CSC representations of matrix $A$
\vspace{2pt} %\Statex
\Function{Partition}{$A$}
  \State \Call{Decompose}{$[0,\;\mathrm{nrows}) \times [0,\;\mathrm{ncols})$}
  \State Discard all blocks $B$ with $\mathrm{width}(B) = 1$
  \If{$\sum_B \mathrm{nnz}(B) < 0.08 \times \mathrm{nnz}(A)$} reject $A$ \EndIf
  \State \Return accepted blocks
\EndFunction
\vspace{2pt} %\Statex
\Function{Decompose}{$R = [r_0, r_1) \times [c_0, c_1)$}
  \If{$\mathrm{area}(R) < A_{\min}$} \Return \EndIf
  \State $B \gets \Call{FindBlock}{A_{\textsc{csr}}, R}$
  \State $B' \gets \Call{FindBlock}{A_{\textsc{csc}}, R}$
  \If{$\mathrm{score}(B') > \mathrm{score}(B)$} $B \gets B'$ \EndIf
  \If{$\mathrm{area}(B) < A_{\min}$} \Return \EndIf
  \If{any row or column of $B$ is already used} \Return \EndIf
  \State \Call{Trim}{$B$} \Comment{Remove low-density edges}
  \If{$\mathrm{density}(B) < 0.5$ \textbf{or} $\mathrm{area}(B) < A_{\min}$} \Return \EndIf
  \State Accept $B$; mark its rows and columns as used
  \State \Call{Decompose}{$[r_0,\; B.r_0) \times [c_0,\; c_1)$} \Comment{Above}
  \State \Call{Decompose}{$[B.r_1,\; r_1) \times [c_0,\; c_1)$} \Comment{Below}
  \State \Call{Decompose}{$[B.r_0,\; B.r_1) \times [c_0,\; B.c_0)$} \Comment{Left}
  \State \Call{Decompose}{$[B.r_0,\; B.r_1) \times [B.c_1,\; c_1)$} \Comment{Right}
\EndFunction
\end{algorithmic}
\end{algorithm}

Algorithm~\ref{alg:partitioner} shows the overall partitioning procedure.
The algorithm takes the matrix in both CSR and CSC representations and recursively decomposes it into \ddense blocks.
For each rectangular region, the block search subroutine (Algorithm~\ref{alg:findblock}) is invoked on both representations --- some blocks are discovered more naturally by iterating along rows than columns --- and the higher-scoring result is kept.
Once a valid block is found, its rows and columns are marked as \emph{used} and the algorithm recurses on the sub-regions above, below, left, and right of the accepted block.
Algorithm~\ref{alg:partitioner} maintains the invariant that no two claimed blocks overlap, ensuring a partitioning compatible with the VBR format.
Before acceptance, a trimming step iteratively removes boundary rows or columns whose deletion improves the block's overall density, evaluating all four edges until no single-edge removal helps.
Any discovered blocks consisting of a single column are discarded, as they cannot benefit from dense execution.
Finally, the matrix is deemed suitable for \sys{}'s code generation only if at least 8\% of its nonzeros are covered by accepted blocks; matrices below this threshold lack sufficient dense structure to justify VBR conversion.
The constants $0.5$ and $0.08$ in Algorithm~\ref{alg:partitioner} instantiate the density floor~$\delta$ and the coverage cutoff at their empirically chosen defaults.
Lowering $\delta$ admits blocks whose padded zeros cost the dense kernels more than dense execution saves, while raising it rejects blocks that are profitable to execute densely.
Similarly, lowering the coverage cutoff admits matrices whose blocks cover too few non-zeros to offset the overhead of dense execution, while raising it excludes matrices that would benefit.\footnote{\autoref{sec:threshold} evaluates the trade-offs of varying this cutoff.}

The core dense block search subroutine is shown in Algorithm~\ref{alg:findblock}.
For a given region and matrix representation (CSR or CSC), the algorithm sweeps over all starting positions along the \emph{outer} dimension---rows for CSR, columns for CSC.
For each starting position~$o_0$, it incrementally extends the outer span by advancing~$o_1$, accumulating a histogram~$H$ that records, for each inner-dimension index, the count of nonzeros across the current span~$[o_0, o_1]$.
A sliding window over~$H$ then identifies the inner-dimension range that maximizes the block score: when the window's density falls below~0.5, the left pointer advances to shrink the window; when the area constraint is not yet met, the right pointer advances to expand it.
This scan is amortized linear in the inner dimension, so the total work per starting position is dominated by the cost of histogram construction.

\begin{algorithm}[t]
\caption{Dense Block Search}\label{alg:findblock}
\begin{algorithmic}[1]
\Require Compressed matrix $M$ (CSR or CSC); region $R$
\vspace{2pt} %\Statex
\Function{FindBlock}{$M, R$}
  \State $B.\mathit{score} \gets -1$; $s_{\min} \gets \lceil\sqrt{A_{\min}}\,\rceil$
  \ParFor{each outer start $o_0$ in $R$}
    \State $H[0 \ldots L_{\mathrm{in}}{-}1] \gets 0$ \Comment{Thread-local histogram}
    \For{$o_1 \gets o_0$ \textbf{to} outer end of $R - 1$}
      \For{each nnz at inner index $j$ in slice $o_1 \cap R$}
        \State $H[j] \gets H[j] + 1$
      \EndFor
      \State $s \gets o_1 - o_0 + 1$
      \If{$s < s_{\min}$} \textbf{continue} \EndIf
      \State $\mathit{nnz} \gets 0$;\; $i_0 \gets 0$
      \For{$i_1 \gets 0$ \textbf{to} $L_{\mathrm{in}} - 1$}
        \State $\mathit{nnz} \gets \mathit{nnz} + H[i_1]$
        \While{$i_0 \leq i_1$}
          \State $a \gets s \times (i_1 - i_0 + 1)$ \Comment{Block area}
          \If{$a < A_{\min}$} \textbf{break} \EndIf
          \State $d \gets \mathit{nnz}\, /\, a$ \Comment{Block density}
          \If{$d \geq 0.5$}
            \State $\mathit{sc} \gets a \cdot (d - 0.5)^{1.5}$
            \If{$\mathit{sc} > B.\mathit{score}$} update $B$ \EndIf
            \State \textbf{break}
          \EndIf
          \State $\mathit{nnz} \gets \mathit{nnz} - H[i_0]$;\; $i_0 \gets i_0 + 1$
        \EndWhile
      \EndFor
    \EndFor
  \EndParFor \Comment{Merge thread-local results}
  \State \Return $B$
\EndFunction
\end{algorithmic}
\end{algorithm}

% \paragraph{Parallelization.}
% \sys{} parallelizes the outer loop of Algorithm~\ref{alg:findblock} using OpenMP with dynamic scheduling.
% Each thread maintains a private histogram and thread-local best candidate; after all iterations complete, thread-local results are merged via a critical section to determine the global best.

\subsection{Band Extractor}
\label{sec:band_partitioner}

\lstinline{BandExtractor} identifies diagonal band regions in a sparse matrix and returns them in VDIA format.
Where \lstinline{BlockExtractor} looks for rectangular dense blocks, \lstinline{BandExtractor} looks for groups of diagonals that are densely populated over contiguous row ranges.
Conceptually, the extraction algorithm mirrors the block extractor: it uses a greedy, density-and-area-thresholded search to identify diagonal ribbons rather than rectangular regions, retaining a region only if it is \ddense{}---governed by the same density floor~$\delta$ used for blocks.

Diagonal band extraction proceeds in two stages.\footnote{The full band extraction and segment-growth algorithms are given in \autoref{sec:bandextractor_algorithms}.}
Candidate diagonal offsets are ranked by nonzero count and processed greedily.
For each diagonal~$d$, a \textsc{FitRibbon} step determines how many diagonals above and below~$d$ can be included at each row while keeping density above~$\delta$, yielding a lower and upper extent~$(l_i, u_i)$ per row.
A \textsc{GrowSegments} step then assembles these per-row extents into contiguous segments: starting from each row that has nonzero extent, it greedily extends a segment row by row while maintaining a rolling-maximum extent~$(L, U)$, recomputing the segment's nonzero count exactly whenever the extent widens and accumulating it otherwise.
A segment is terminated when its density drops below~$\delta$, and only segments spanning at least~$\ell_{\mathrm{seg}}$ rows are retained.
A region is accepted only if its aggregate area exceeds~$A_{\min}$ and its density exceeds~$\delta$; accepted rows are masked out, preventing overlapping rediscoveries at neighbouring offsets.
The surviving regions are stored in VDIA format described below.
% : per-segment metadata (row start, row count, number of diagonals, pointers into the value and diagonal-index arrays) plus values stored in diagonal-major order
%The full extraction and segment-growth algorithms are given in \autoref{sec:bandextractor_algorithms}.

\subsubsection{The VDIA Format}
\label{sec:vdia_format}

Many sparse matrices exhibit a diagonal band structure of varying size across different row ranges.
Existing representations for diagonal bands do not capture this variability: DIA~\cite{saad2003iterative} assumes that every stored diagonal spans the full matrix, padding shorter bands with explicit zeros, while HDIA~\cite{filippone2017spmv} reduces this padding by splitting the matrix into row groups, but fixes the groups to a uniform height chosen independently of the matrix's structure.
To address this, we have developed a novel representation, called VDIA, which relaxes these restrictions: bands may begin and end at arbitrary rows, and different row ranges may carry different numbers of diagonals at different offsets.
VDIA partitions the banded structure into \emph{segments}, each covering a contiguous range of rows with its own set of diagonals.
A segment is described by its starting row, row count, number of diagonals, and pointers into shared value and diagonal-index arrays.
Values within a segment are stored in diagonal-major order: for each diagonal offset, the values for all rows in the segment are laid out contiguously, enabling vectorized access along diagonals.
We illustrate the format in Fig.~\ref{fig:vdia}: Segment~0 covers rows~1--4 with 3~diagonals along the main diagonal, while Segment~1 covers rows~6--8 with 4~diagonals centered on offset~$-4$, well below the main diagonal.
Fig.~\ref{fig:vdia_indirection} shows the per-segment metadata and the shared value and diagonal-index arrays.
The \lstinline{BandExtractor} (\autoref{sec:band_partitioner}) analyzes a sparse matrix and produces a VDIA format object that the kernel layer can consume directly.

\renewcommand{\arraystretch}{1.1}
\begin{figure*}
    \centering
    \begin{adjustbox}{minipage=0.38\linewidth}
    \begin{subfigure}{0.99\textwidth}
      \centering
      {\footnotesize
        \begin{tabular}{| c c c c c c c c c c |}
            \hline
            0 & 0 & 0 & 0 & 0 & 0 & 0 & 0 & 0 & 0 \\
            \cellcolor{blue!30}4 & \cellcolor{blue!30}7 & \cellcolor{blue!30}2 & 0 & 0 & 0 & 0 & 0 & 0 & 0 \\%[0.3em]
            0 & \cellcolor{blue!30}1 & \cellcolor{blue!30}5 & \cellcolor{blue!30}3 & 0 & 0 & 0 & 0 & 0 & 0 \\%[0.3em]
            0 & 0 & \cellcolor{blue!30}2 & \cellcolor{blue!30}8 & \cellcolor{blue!30}1 & 0 & 0 & 0 & 0 & 0 \\%[0.3em]
            0 & 0 & 0 & \cellcolor{blue!30}3 & \cellcolor{blue!30}6 & \cellcolor{blue!30}4 & 0 & 0 & 0 & 0 \\%[0.3em]
            0 & 0 & 0 & 0 & 0 & 0 & 0 & 0 & 0 & 0 \\
            \cellcolor{Salmon}3 & \cellcolor{Salmon}2 & \cellcolor{Salmon}5 & \cellcolor{Salmon}1 & 0 & 0 & 0 & 0 & 0 & 0 \\%[0.3em]
            0 & \cellcolor{Salmon}4 & \cellcolor{Salmon}3 & \cellcolor{Salmon}7 & \cellcolor{Salmon}2 & 0 & 0 & 0 & 0 & 0 \\%[0.3em]
            0 & 0 & \cellcolor{Salmon}1 & \cellcolor{Salmon}6 & \cellcolor{Salmon}2 & \cellcolor{Salmon}8 & 0 & 0 & 0 & 0 \\%[0.3em]
            0 & 0 & 0 & 0 & 0 & 0 & 0 & 0 & 0 & 0 \\
            \hline
        \end{tabular}
      }
        \caption{Matrix with two diagonal band segments}
    \end{subfigure}
    \end{adjustbox}
    \begin{adjustbox}{minipage=0.61\linewidth}
    \begin{subfigure}{0.99\textwidth}
        {\small
        \begin{itemize}
            \item \lstinline{val}: [\colorbox{blue!30}{4, 1, 2, 3, 7, 5, 8, 6, 2, 3, 1, 4}, \colorbox{Salmon}{3, 4, 1, 2, 3, 6, 5, 7, 2, 1, 2, 8}]. Values stored in diagonal-major order within each segment.
            \item \lstinline{idiag}: [\colorbox{blue!30}{$-$1, 0, 1}, \colorbox{Salmon}{$-$6, $-$5, $-$4, $-$3}]. Diagonal offsets for each segment.
            \item \lstinline{seg_row_start}: [\colorbox{blue!30}{1}, \colorbox{Salmon}{6}]. Starting row of each segment.
            \item \lstinline{seg_nrows}: [\colorbox{blue!30}{4}, \colorbox{Salmon}{3}]. Number of rows in each segment.
            \item \lstinline{seg_ndiags}: [\colorbox{blue!30}{3}, \colorbox{Salmon}{4}]. Number of diagonals in each segment.
            \item \lstinline{seg_val_ptr}: [\colorbox{blue!30}{0}, \colorbox{Salmon}{12}]. Index into \lstinline{val} where each segment's values begin.
            \item \lstinline{seg_idiag_ptr}: [\colorbox{blue!30}{0}, \colorbox{Salmon}{3}]. Index into \lstinline{idiag} where each segment's offsets begin.
        \end{itemize}
        }
        \caption{Indirection arrays for the VDIA matrix}
        \label{fig:vdia_indirection}
    \end{subfigure}
  \end{adjustbox}
  \vspace{-5pt}
    \caption{Matrix with two diagonal bands represented in the Variable Diagonal format.
      % \colorbox{blue!30}{Segment~0} spans rows~1--4 with 3~diagonals along the main diagonal; \colorbox{Salmon}{Segment~1} spans rows~6--8 with 4~diagonals centered on offset~$-4$.
    }
    \label{fig:vdia}
\end{figure*}

\subsection{Format Composition}
\label{sec:format_composition}

As described in \autoref{sec:plan_interface}, each extractor operates on the residual matrix left by the previous extraction, allowing \sys{} to naturally support multi-format compositions.
To do so, the final extractor must be able to account for any unstructured non-zeros that were not claimed by its predecessors.
In practice we use an extractor that targets the CSR representation, since neither VBR nor VDIA is guaranteed to capture all non-zeros: scattered elements that do not belong to any dense block or diagonal band remain in the residual.
The order of components matters when defining a format compositionally, since each extractor sees only the non-zeros that previous extractors did not claim.
For example, extracting VDIA before VBR removes banded structure first, so the block partitioner operates on a residual that may expose different---or fewer---rectangular blocks than it would find in the original matrix.
The reverse order (VBR before VDIA) would remove rectangular blocks first, potentially fragmenting diagonal bands in the residual.
\sys{} does not prescribe a fixed ordering; users can choose the extraction sequence they believe best matches the structure of the target matrix.

\section{Evaluation}
\label{sec:evaluation}

Our evaluation considers the following key questions:

\begin{itemize}
\item \textbf{RQ1} How do the performance improvements of \sys{}
  compare to that of prior approaches?
\item \textbf{RQ2} Can \sys{} generate hybrid formats capturing a
  variety of structured sparsity patterns and matrix operations?
\item \textbf{RQ3} Does \sys{} scale well with multiple threads? %
\end{itemize}

\paragraph{Evaluation scenario}
Because extraction and code generation in \sys{} depend only on a matrix's sparsity structure, not its values, a generated program runs over any matrix sharing that structure.
Our evaluation thus uses the standard inspector–executor scenario~\cite{inspector93} and reports the runtime of the generated program, since the one-time generation cost is amortized across the many invocations in which values change but structure does not.

% Extraction and code generation in \sys{} are independent of the particular values of the non-zero elements of a matrix: both depend only on its sparsity structure.
% As a consequence, \sys{} generates programs that can operate over different matrices that have the same sparsity structure.
% Our evaluation targets this standard inspector--executor scenario~\cite{inspector93}: we measure the runtime of the generated program, whose one-time generation cost is amortized across the many invocations in which the values of the matrix may change but its sparsity structure does not.

\paragraph{Experimental setup}
All experiments were conducted on a single-socket Intel(R) Xeon(R)
w7-3445 system with 20 physical cores (20 hardware threads), each
running at up to 4.8 GHz. The system has 52.5 MiB of shared L3 cache,
40 MiB of total L2 cache (2 MiB per core), and 48 KB of L1 data cache
per core. Each program was compiled using GCC v11.4.0 with the
\lstinline{-O3} \lstinline{-march=native}
\lstinline{-funroll-all-loops} \lstinline{-mavx}
\lstinline{-mprefer-vector-width=512} optimization flags. For SpMV,
each generated benchmark binary performs 30 timed iterations
internally and is invoked 30 times, yielding 900 samples per
configuration. % ; this higher sample count is necessary because SpMV's
% shorter runtimes make it more susceptible to system noise
We remove outliers using decile filtering and report the mean of the
remaining samples. For SpMM, the dense input has 512 columns; the binary performs 10 internal iterations and is invoked 10 times (100 samples per configuration), as the longer per-invocation runtimes produce more stable measurements.

\paragraph{Benchmark construction}
We evaluate \sys{} using SuiteSparse, a publicly
available set of sparse matrices drawn from real-world applications~\cite{SuiteSparse}.

\begin{itemize}
\item \textbf{VBR:} Applying our block partitioner (\autoref{sec:block_partitioner}) to 1578 SuiteSparse matrices identified 117 matrices with \ddense{} blocks. The 55 of these in which at least 8\% non-zeros are covered by \ddense{} blocks are used in our VBR+CSR evaluation.
  These matrices range from 18{,}552 to 16.2 million non-zeros and have shapes that range from small and square to highly rectangular (up to $512 \times 394{,}792$).
Per-matrix dimensions, block counts, densities, and coverage appear in Appendix \ref{sec:vbr_csr_detailed}.
% The number of \ddense{} blocks identified ranges from 1 to 170, with average block densities between 56\% and 100\%.
% Block coverage---the fraction of non-zeros residing in \ddense{} blocks---varies widely, from 8\% to 99\%.
\item \textbf{VDIA:} We applied our band extractor (\autoref{sec:band_partitioner}) to the 117 matrices in which the block partitioner identified \ddense{} blocks.\footnote{We restricted the band extractor to the matrices already selected by the block partitioner in order to enable the VDIA+VBR+CSR comparison below.} The band extractor identified 24 matrices with diagonal band structure suitable for VDIA extraction; these are used in our VDIA+CSR and VDIA+VBR+CSR evaluations.
\citet{fukaya2021mhdc} evaluate 20 matrices with diagonal bands---out of which we found 4 matrices that were valid for our use case (after filtering out matrices with more than 23M non-zeros), and also evaluated over them in the VDIA+CSR configuration.
\end{itemize}

\paragraph{Kernels}
We evaluate \sys{} on two widely used sparse matrix operations: Sparse
Matrix-Vector multiplication (SpMV),
$y_i = \sum_{j} A_{ij} \cdot x_j$, and Sparse Matrix-Matrix
multiplication (SpMM), $C = A \cdot B$ with $B$ dense, which play
critical roles in numerous domains, including iterative
solvers~\cite{saad2003iterative, sellp,CSB}, graph
processing~\cite{page1999pagerank, ligra}, scientific
computing~\cite{csr5,Im:CSD-00-1104,sadayintro,1dvbl}, and machine
learning~\cite{featgraph,dgl}.  The associativity of the $+$ operator
lets us demonstrate the system without having to solve the orthogonal
data-dependency problem.
%To demonstrate the practical applicability of our approach, we focus on evaluating the SpMV and SpMM operations, due to the commutative and associative nature of the $+$ operator, which allows us to demonstrate the capabilities of our system without having to solve an orthogonal data dependency problem.

\paragraph{Baselines}
We compare \sys{} against five baselines:
\begin{itemize}[leftmargin=*]
\item \textbf{UZP}~\cite{uzp} \textit{(SpMV)}: An inspector--executor
  system that represents structured regions as sets of polyhedra and
  leaves irregular elements in CSR/COO.  This is a natural hybrid
  format against which to test \sys{}'s \ddense{} blocks.
\item \textbf{MKL}~\cite{mkl} \textit{(SpMV+SpMM)}: The
  vendor-optimized sparse BLAS kernel from Intel's oneAPI Math
  Kernel Library. %
\item \textbf{SpV8}~\cite{spv8} \textit{(SpMV)}:
  A state-of-the-art vectorized SpMV kernel using AVX-512 gather for branchless column access.
  Its branchless design makes runtime scale near-linearly with the number of non-zeros, making it an especially informative comparison point for \sys{}, whose gains depend on the sparse dispatch shrinking proportionally as non-zeros move into \ddense{} blocks.
  % A CSR-based SpMV
  % implementation that uses AVX-512 gather instructions for vectorized
  % column access. Removing non-zero elements from a matrix stored in
  % CSR does not always translate to a proportional SpMV speedup due to
  % effects such as branch mispredictions; SpV8's branchless vectorized
  % design achieves close to expected linear relationship. This makes
  % SpV8 a particularly informative comparison point for our approach,
  % which relies on the sparse dispatch becoming proportionally faster
  % as non-zeros are moved into \ddense{} blocks.
  % SpV8 is a new format, which lets us compare \sys against a state-of-the-art non-CSR format.
\item \textbf{Sparse Register Tiling (SpReg)}~\cite{registertiling}
  \textit{(SpMM)}: A code generator that mines for submatrices with
  similar sparsity patterns and constructs regular loops over them
  using an unroll-and-sparse-jam technique. %
\item \textbf{Naive} \textit{(SpMV+SpMM)}: A textbook
  CSR-based kernel. %
\end{itemize}
We do not include techniques~\cite{piecewise, kazemcodelet,
  sparsepolyhedral} that were previously shown to perform worse than
one of our baselines.

\paragraph{Implementation}
% \sys{}'s block partitioner requires the \ddense{} blocks of a matrix to
% account for at least 8\% of its total non-zeros, as below this
% threshold, the overhead of dense execution tends to outweigh its
% performance benefits.
In our evaluation, \sys{} dispatches \ddense{} blocks to either
optimized BLAS routines (\eg \texttt{cblas\_dgemv} for SpMV,
\texttt{cblas\_dgemm} for SpMM) or shape-specialized handwritten
kernels, based on the dimensions of the block.
VDIA segments are dispatched to a diagonal kernel that iterates over diagonals within each segment, exploiting the regular stride-one access pattern for vectorization.
Our band extractor accepts only \ddense{} segments; for the VDIA evaluation we set $\delta = 75\%$, so every extracted band is at least $75\%$ dense.\footnote{The value was chosen empirically: extraction with lower floors (\eg $\delta = 50\%$) yielded slightly lower overall speedups.}
In all configurations, the residual non-zeros that remain after extraction are stored in CSR and dispatched to one of the sparse baselines described above.

\subsection{\sys{} Performance (RQ1+RQ2)}
\label{sec:rq1}

To compare our approach to existing techniques for structured sparse
matrix computation, we evaluate on three plans capturing three
distinct sparsity structures: namely \textbf{VBR+CSR},
\textbf{VDIA+CSR}, and \textbf{VDIA+VBR+CSR}.
% \begin{itemize}[leftmargin=*]
% \item[$-$] \textbf{VBR+CSR}: extract rectangular dense blocks, dispatch the residual to CSR.
% \item[$-$] \textbf{VDIA+CSR}: extract diagonal band segments, dispatch the residual to CSR.
% \item[$-$] \textbf{VDIA+VBR+CSR}: extract diagonal bands first, then rectangular blocks from the remaining non-zeros, dispatch the final residual to CSR.
% \end{itemize}

\paragraph{VBR+CSR vs CSR}
\label{sec:vbr_csr_eval}

\begin{figure*}[t!]
    \centering
    \includegraphics[width=\textwidth]{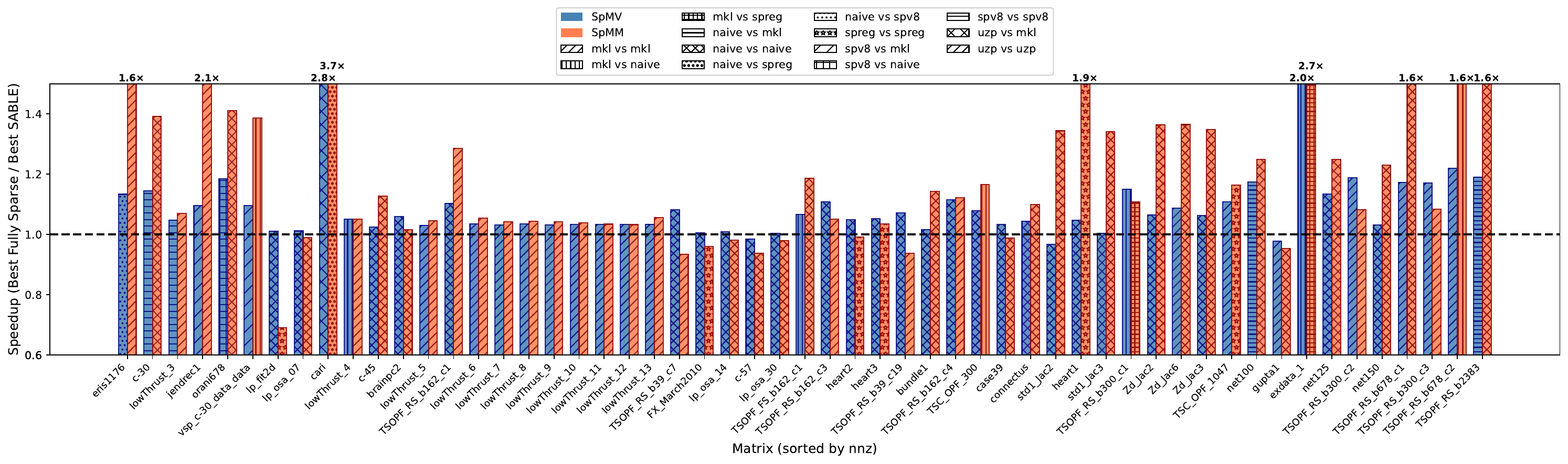}
       \vspace{-20pt}
    \caption{Speedup of \sys{} using VBR+CSR over the best
      fully-sparse baseline for SpMV and SpMM. Each bar's hatching
      indicates \sys{}'s best sparse dispatch and the best
      baseline. Matrices with a speedup of $\ge 1.5\times$ are
      labeled with their value.}
    \label{fig:bench_executor}
    \vspace{-10pt}
\end{figure*}

We applied \sys{} with block extraction to the 55 matrices in our
benchmark suite.
\autoref{fig:bench_executor} reports the single-threaded speedup of the best \sys{} configuration over the best fully-sparse baseline for each matrix.\footnote{A full per-matrix breakdown---dimensions, block counts, densities,
coverage, and the time and speedup against every baseline for all 55
matrices---is given in Appendix \ref{sec:vbr_csr_detailed}.}
Here, the best \sys{} configuration is the fastest combination of \sys's dense dispatch for \ddense{} blocks and a sparse library for the remaining values, while the best fully-sparse baseline is the fastest baseline applied to the entire matrix.

For SpMV, \sys{} achieves a geometric mean speedup of $1.10\times$
over the best fully-sparse baseline, with speedups on 52 out of 55
matrices. The largest SpMV gain is on \textit{exdata\_1}
($2.01\times$), where 99.3\% of non-zeros reside in a single dense
block, allowing \sys{} to dispatch nearly all computation to
\texttt{cblas\_dgemv}.
The magnitude of the gain is governed primarily by block coverage: matrices with high coverage show substantial gains, while matrices with coverage closer to the 8\% acceptance threshold typically still achieve modest speedups ($1.01$--$1.05\times$) by accelerating just the dense portion.
The three SpMV slowdowns ($0.96$--$0.99\times$) occur on matrices whose blocks have relatively low density, so the dense kernel spends computation on zeros that the savings from shrinking the sparse problem do not compensate; a detailed analysis of these outliers is given in Appendix \ref{sec:vbr_csr_detailed}.

For SpMM, \sys{} achieves a geometric mean speedup of $1.20\times$,
with gains on 44 out of 55 matrices. SpMM generally shows larger
speedups than SpMV, as the dense kernel benefits from greater data
reuse across the columns of the output matrix.
SpMM also benefits more than SpMV from the cache locality of dense block storage, allowing some matrices with modest block coverage ($11$--$15\%$) to sustain speedups of $1.31$--$1.36\times$.
The 11 SpMM slowdowns occur where the extracted blocks do not offset the overhead of dense dispatch at their coverage levels; in the worst cases, the residual sparse dispatch also fails to become proportionally faster as non-zeros are extracted.

\paragraph{VDIA+CSR vs CSR}
\label{sec:vdia_csr_eval}

\begin{figure}[t!]
    \centering
    \includegraphics[width=\columnwidth]{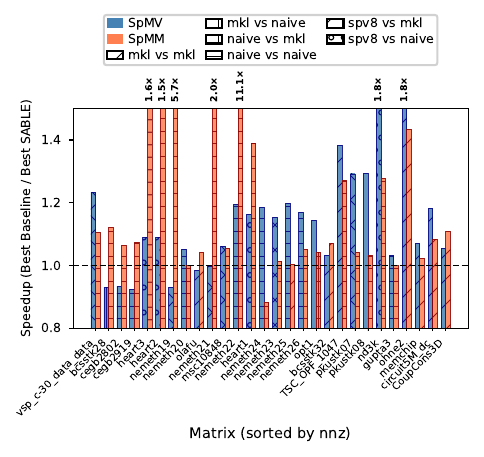}
   \vspace{-20pt}
    \caption{Speedup of \sys{} using VDIA+CSR over the best
      fully-sparse baseline for SpMV and
      SpMM. Clipped-bar labels are as in
      \autoref{fig:bench_executor}.}
    \label{fig:vdia_csr_best}
       \vspace{-10pt}
\end{figure}

We next evaluate \sys{}'s diagonal band extraction by adding VDIA
to a plain CSR configuration.
\autoref{fig:vdia_csr_best} shows the speedup of the best VDIA+CSR configuration over the best CSR baseline for each matrix.\footnote{The full per-matrix times and speedups against every backend are reported in
Appendix \ref{sec:vdia_csr_detailed}.}
%In the interest of time, we did not evaluate SpReg (SpMM) or UZP (SpMV) in the VDIA
%configurations---neither as fully-sparse baselines nor as \sys{}'s residual CSR
%dispatch; they appear only in the VBR+CSR evaluation (\autoref{sec:vbr_csr_eval}).}

As with VBR+CSR, the gains differ sharply between SpMV and SpMM, with SpMM benefiting more than SpMV.
Because our band extractor accepts only \ddense{} segments --- here with
$\delta = 75\%$ --- every extracted band has roughly the same density, so band
coverage (the fraction of non-zeros the diagonal kernel captures) is what differentiates the matrices.
The diagonal kernel loads each band element once and reuses it across every
column of the dense matrix, so its regular stride-one access pattern
amortizes the cost of dispatching a separate kernel and exposes data
reuse.
SpMV, by contrast, performs a single multiply-add
per element.

For SpMM, \sys{} speeds up 26 of 28 matrices (geometric mean $1.31\times$),
and the speedup tracks band coverage closely: gains reach $11.14\times$ for the highest-coverage matrices and fall off smoothly to near-neutral for matrices whose bands cover under $1\%$ of non-zeros.
For SpMV, \sys{} speeds up 22 of 28 matrices (geometric mean $1.14\times$),
but the gains do not track band coverage: lacking SpMM's per-element reuse, the diagonal kernel's only advantage over CSR is the improved cache behavior of its regular stride-one accesses, a benefit that is small but largely independent of coverage.
%\TODO{This explanation cannot account for a 15--20\% whole-program gain when over 99\% of the work is the residual CSR pass}
The six slowdowns are matrices where this benefit is too small to offset the cost of splitting the computation into a diagonal pass and a residual CSR pass. % (per-matrix analysis in \autoref{sec:vdia_csr_detailed}).

\paragraph{VDIA+VBR+CSR vs VBR+CSR}
\label{sec:vdia_vbr_csr_eval}

\begin{figure}[t!]
    \centering
    \includegraphics[width=\columnwidth]{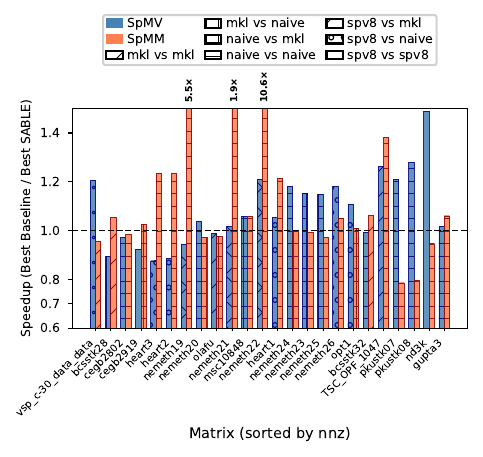}
    \vspace{-20pt}
    \caption{Speedup of \sys{} using VDIA+VBR+CSR over
      the best VBR+CSR baseline for SpMV and SpMM. Clipped-bar labels
      are as in \autoref{fig:bench_executor}.}
    \label{fig:vdia_vbr_csr_best}
    \vspace{-10pt}
\end{figure}

Finally, we evaluate the full three-format composition.
\autoref{fig:vdia_vbr_csr_best} shows that adding VDIA on top of
VBR+CSR yields a geometric mean speedup of $1.08\times$ for SpMV, with
gains on 16 of 24 matrices, and $1.25\times$ for SpMM, with gains on
14 of 24 matrices, over the best VBR+CSR baseline. \footnote{The full
  per-matrix times and speedups---with VBR+CSR rather than plain CSR
  as the baseline for each backend---are reported in
  \autoref{table:vdia_vbr_csr} in the appendix
  (\autoref{sec:vdia_vbr_csr_detailed}).} The incremental benefit is
smaller than adding VDIA to plain CSR (\autoref{fig:vdia_csr_best}),
because VBR+CSR has already extracted the matrix's \ddense{} blocks and
dispatched them to efficient kernels, raising the bar that the
diagonal extraction must clear. As before, the SpMM gains track band
coverage closely.

Overall, composing VDIA with VBR+CSR helps most for matrices with
strong banded structure that VBR does not already capture, and, for
similar reasons as before, the SpMV and SpMM outcomes can diverge in
either direction depending on which baseline kernel the band must
improve upon.  A per-matrix analysis of these divergences is given in
the appendix.

\paragraph{Discussion}
These results confirm that \sys{} improves performance over prior
approaches on real-world matrices for both SpMV and SpMM across all
three format configurations, % : geometric-mean speedups of
% $1.10\times$/$1.20\times$ for VBR+CSR (with gains on 52/55 and 44/55
% matrices) and $1.14\times$/$1.32\times$ for VDIA+CSR (23/29 and 27/29)
% over the best fully-sparse baselines, and an incremental
% $1.09\times$/$1.22\times$ for VDIA+VBR+CSR over the VBR+CSR baseline
% (17/25 and 14/25),
with the magnitude of improvement correlating
primarily with the fraction of non-zeros captured in structured
regions.
Composing formats---adding VDIA on top of VBR+CSR---yields further
gains for matrices with banded structure, confirming that the formats
capture complementary dense substructure \textbf{(RQ1)}.
Notably, all three plans---including two hybrid formats that exist only as
compositions of independently defined extractors and kernels---required
no changes to the framework \textbf{(RQ2)}.
%A detailed per-matrix analysis of all three configurations---including the outliers noted above---is given in \autoref{sec:detailed_results} in the appendix.

\subsection{Parallelization (RQ3)}
\label{sec:rq2}

To understand how \sys{} scales when parallelized, we measure the speedup of \sys{} over the best fully-sparse baseline at multiple thread counts for both SpMV and SpMM.\footnote{We exclude SpReg from this comparison, as we were unable to integrate its parallelization with our system.}
We restrict the parallel evaluation to the VBR+CSR configuration, since \sys{}'s parallel speedups derive primarily from dispatching \ddense{} blocks to multithreaded BLAS; the VDIA kernels are not yet integrated with our multithreaded execution, so we leave their parallelization to future work.

\begin{figure}[t!]
    \centering
    \includegraphics[width=\columnwidth]{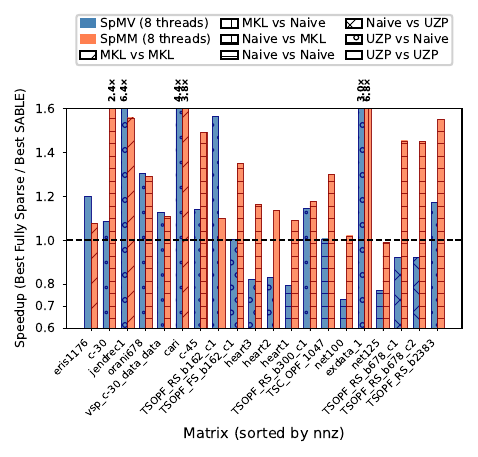}
    \vspace{-20pt}
    \caption{Speedup of \sys{} using VBR+CSR over the best
      fully-sparse baseline at 8 threads on the 20 matrices with
      ${\geq}15\%$ block coverage, block aspect ratio ${\leq}100$.
      Hatching and clipped-bar labels as in
      \autoref{fig:bench_executor}.}
    \label{fig:parallel_spmv}
    \vspace{-10pt}
\end{figure}

\sys{} parallelizes each dispatch with the mechanism best suited to
it.  \ddense{} blocks dispatched to a BLAS routine (\eg
\texttt{cblas\_dgemm}, MKL) rely on the library's built-in
multithreading, whereas blocks dispatched to \sys{}'s handwritten
shape-specialized kernels are parallelized with OpenMP.  The sparse
residual dispatch brings its own parallelization (\eg MKL sparse,
UZP).  Not all matrices that benefit from \sys{} in the
single-threaded setting are good candidates for parallelization.
Matrices whose \ddense{} blocks cover fewer than 15\% of non-zeros do
not offset threading overhead, and matrices with extremely elongated
blocks cannot be parallelized efficiently by the BLAS kernel.
Filtering these cases reduces the evaluation set from 55 to 20
matrices.\footnote{Appendix \ref{sec:par_threshold} reports how each
  filter affects parallel performance in detail.}
% (\autoref{table:par_coverage} and \autoref{table:par_aspect}).}
\autoref{fig:parallel_spmv} shows the per-matrix speedup of \sys{}
over the best fully-sparse baseline at 8 threads for both SpMV and
SpMM; we choose 8 because this is the thread count at which the
baselines we evaluated on achieve their best average scaling across
our benchmark matrices.

For SpMV, \sys{} achieves speedups on 13 of 20 matrices
(geometric mean $1.26{\times}$), with the largest gains on
\textit{jendrec1} ($6.4\times$) and \textit{cari}
($4.4\times$).
The seven slowdowns fall into two groups: matrices with only moderate block coverage, where the sparse residual dominates total runtime and the fully-sparse baselines parallelize it competitively, and matrices with many small blocks, whose per-block GEMV work is too small to amortize the threading overhead of a parallel BLAS call.
The higher SpMV loss rate is expected, since SpMV is
bandwidth-bound and parallel GEMV gains less from multithreading than
GEMM.% , and the fully-sparse baselines' own thread-level parallelism is
% often competitive for the sparse portion that dominates total SpMV time.

For SpMM, \sys{} achieves speedups on 19 of 20 matrices (geometric
mean $1.48{\times}$), with the largest gains on \textit{exdata\_1}
($6.8\times$) and \textit{cari} ($3.8\times$).  A per-matrix breakdown
of the SpMV slowdowns, together with a detailed analysis of the thread
scaling behavior of all \sys{} sparse dispatches and fully-sparse
baselines on four representative matrices, is given in
\autoref{sec:thread_scaling}.  These results confirm that \sys{}'s
parallel advantage is strongest when the identified \ddense{} blocks
constitute a significant fraction of total non-zeros \textbf{(RQ3)}.

\section{Related Work}
\label{sec:related_work}

\paragraph*{Sparse matrix representations}
CSB~\cite{CSB}, BCSR~\cite{bcsr}, and SELL-P~\cite{sellp} are fixed-size block formats.
On the diagonal side, HDIA~\cite{filippone2017spmv} reduces DIA's padding by splitting the matrix into equally sized row groups; VDIA's segments are instead placed and sized adaptively by \sys{}'s density analysis.
\citet{UBCSR} extend BCSR~\cite{DBLP:journals/ijhpca/ImYV04} to reduce fixed-size blocking's fill-in, lowering indexing overhead and enabling register-level tiling.
Our approach builds on these techniques.
TACO~\cite{taco, tacoiter} and UniSparse~\cite{unisparse} provide abstractions over storage formats and the iteration spaces that traverse them, but cannot represent variable-sized blocks.
The MLIR sparse tensor dialect~\cite{Bik_2022} brings TACO-style per-tensor format annotations and sparse code generation to MLIR.
Finch~\cite{finch} extends TACO-style level abstractions with control flow and a user-extensible vocabulary of structures (\eg runs and bands), but like TACO it expresses structure through the choice of representation rather than discovering it in a particular matrix.
SparseTIR~\cite{sparsetir} provides a TACO-like language to specify operations over matrices as a composition of different storage formats based on loop index constructs.
AlphaSparse~\cite{alphasparse} goes further still and generates a new format and SpMV kernel for each input matrix, but its hardcoded operators only partition matrices into row or column bands and fixed-size blocks, so it cannot detect variable-sized blocks from local density structure. %; it also generates GPU kernels, whereas \sys{} targets CPUs.
Mosaic~\cite{mosaic} binds tensor algebra sub-expressions to external library functions through a scheduling API; composition happens at the expression level, whereas \sys{} inspects a concrete matrix, splits that one operand into regions, and binds a kernel to each region.

\paragraph*{Hybrid storage formats}
Several systems compose multiple storage formats to handle matrices containing heterogeneous substructures.
The classic HYB format~\cite{bell2009implementing} splits a matrix between ELL for the regular portion of each row and COO for the irregular remainder.
LAV~\cite{lav} stores the dense region of a matrix in a new format, and the sparse region as CSR.
TileSpMV~\cite{tilespmv} and BDMP-PBP~\cite{DBLP:journals/tjs/ZhaoZWHYF24} divide a matrix into fixed-size blocks and pick the best format for each block.
STile~\cite{stile} finds sub-matrices and appropriate storage formats for each, storing the result as a composition of formats in SparseTIR's IR.
UZP~\cite{uzp} represents a matrix's dense subregions as equations and the rest as COO or CSR.
M-HDC~\cite{fukaya2021mhdc} pairs DIA with CSR; in \sys{}, the analogous split is simply one user-defined plan (VDIA+CSR).
Unlike these systems, whose sets of formats and kernels are fixed, \sys{} treats formats, extractors, and kernels as user-defined components that compose without changes to the framework.

\paragraph*{Re-ordering matrices for better performance}
Other work re-orders matrices for better performance~\cite{cacheoblivious, aspt}.
SpV8~\cite{spv8} groups rows of similar density together and dispatches them to kernels with different vectorization strategies, avoiding branch mispredictions; VNEC~\cite{VNEC} builds on it with a storage format that chunks rows and reduces empty columns to coalesce elements together.
\sys{} does not re-order the input matrix: its extractors identify structure that is already present, such as \ddense{} blocks and diagonal bands, and carve it into the appropriate representation.
In the future, re-ordering could serve as a preprocessing step that exposes such structure for \sys{}'s extractors to capture.

\paragraph*{Inspector-executor}
The inspector--executor model~\cite{inspector93} for sparse matrix operations~\cite{inspectorsparse} is well studied, particularly for SpMV: these approaches generate code specialized to a matrix's sparsity pattern that can be reused across matrices sharing it.
The most closely related are LCM I/E~\cite{kazemcodelet}, which mines partially strided codelets from the matrix's memory accesses to vectorize sparse computations; Sparse Register Tiling~\cite{registertiling}, which uses an unroll-and-sparse-jam technique to achieve register reuse without excessive fill-in; and \citet{piecewise}, which builds regular polyhedra from a trace of the memory access patterns, though its inspection time is too high for larger matrices and it only handles regularly strided indices.
MACVETH~\cite{piecewiseimpact} extends this line of work by performing a peephole optimization over \textit{gather} instructions to facilitate vectorization.
Other widely used inspector--executor systems for sparse tensors include the Sparse Polyhedral Framework~\cite{sparsepolyhedral,spfset, spfpar}, Sympiler~\cite{sympiler}, and ParSy~\cite{parsy}.
In the sparse tensor domain, metaprogramming techniques~\cite{lmspopl, leoarray} have generated code specialized to specific operations, but none compose arbitrary user-defined formats with arbitrary user-defined kernels as \sys{} does.
\citet{dehame:hal-04358187} partially evaluate parts of sparse tensor operations, but are limited to hardcoded rules and do not generalize to arbitrary formats or operations.

\section{Conclusion}
\label{sec:conclusion}

This paper presented \sys{}, a framework for structured sparse matrix computation in which hybrid formats are assembled by users rather than fixed in advance.
Under its plan--extract--dispatch interface, extractors claim the subregions of a matrix that match their target structure, kernels emit specialized C code for the chosen representations, and \sys{} weaves these pieces into a single program.
Our compositional approach is both practical---VDIA, our novel variable-length diagonal format, and the two hybrid formats built from it required no changes to the framework---and profitable, yielding geometric-mean speedups of $1.10$--$1.31\times$ over the best fully-sparse baselines across SpMV and SpMM on real-world SuiteSparse matrices.
%Future work includes parallelizing the VDIA diagonal kernels (\autoref{sec:rq2}), applying re-ordering as a preprocessing step to expose structure that existing extractors can capture (\autoref{sec:related_work}), and growing the library of extractors, kernels, and operations.

\bibliographystyle{ACM-Reference-Format}
\bibliography{sigproc}

% Record the page on which the appendix starts so the Makefile can split
% paper.pdf into only_paper.pdf and appendix.pdf.
\clearpage
\immediate\openout\appendixstartfile=\jobname.appendixstart
\immediate\write\appendixstartfile{\thepage}
\immediate\closeout\appendixstartfile

\clearpage
\appendix

\section*{Appendix}

\section{Instantiating Core Abstractions}%
\label{sec:inst-abstractions}

\subsection{Extractor Example: CSR}%
\label{sec:inst-extractor}

Extractor is any Python object with:
\begin{itemize}[leftmargin=*]
\item A \lstinline{produces} class attribute naming the \lstinline{Format} subclass it returns.
\item An \lstinline{extract(self, A: ResidualMatrix)} method that returns a \lstinline{(Format, ResidualMatrix)} pair.
\end{itemize}

The simplest extractor is \lstinline{CSRExtractor}, which claims all remaining non-zeros as CSR:

\begin{lstlisting}[language=Python]
class CSRExtractor:
    produces = CSR

    def extract(self, A):
        csr = A.to_csr()
        fmt = CSR(
            nrows=A.nrows, ncols=A.ncols,
            nnz=int(csr.nnz),
            indptr=Rep(csr.indptr.tolist(),
                       label="csr_indptr"),
            indices=Rep(csr.indices.tolist(),
                        label="csr_indices"),
            values=Rep(csr.data.tolist(),
                       label="csr_val"))
        return fmt, A.empty_residual()
\end{lstlisting}

Note how the CSR extractor wraps \lstinline{indptr} and \lstinline{indices} in \lstinline{Rep}s because CSR kernels traverse them at run time.
Our VBR extractor, instead, keeps its block metadata as plain lists: kernels consume the block boundaries while generating code, so blocks appear in the emitted program only as constant loop bounds and array offsets, and only the \lstinline{val} array crosses into the program.

\subsection{Kernel Example: naive CSR SpMM loop}%
\label{sec:inst-kernel}

A kernel is any Python object with:
\begin{itemize}[leftmargin=*]
\item An \lstinline{accepts} attribute naming the \lstinline{Format} subclass it operates on.
\item An \lstinline{operation} attribute (inherited from \lstinline{SpmvKernel} or \lstinline{SpmmKernel}).
\item An \lstinline{emit_call(self, fmt, y, x, rhs)} method that returns a C code string (or list of strings) for the dispatch.
\end{itemize}

\begin{lstlisting}[language=Python]
class NaiveCSRSpmm:
    accepts = CSR
    operation = Operation.SPMM

    def emit_call(self, fmt, y, x, rhs):
        nrhs = rhs.shape[1]
        return f"""
for (int i = 0; i < {fmt.nrows}; i++) {{
  for (int p = {fmt.indptr}[i];
       p < {fmt.indptr}[i+1]; p++) {{
    int col = {fmt.indices}[p];
    double a = {fmt.values}[p];
    for (int j = 0; j < {nrhs}; j++) {{
      {y}[i*{nrhs}+j] += a*{x}[col*{nrhs}+j];
    }}
  }}
}}"""
\end{lstlisting}

The kernel author writes a Python f-string that generates C code, and the format's fields resolve according to their class (\autoref{sec:formats}): \lstinline{fmt.nrows} evaluates to a Python integer and is inlined as a constant, while \lstinline{fmt.indptr} is a \lstinline{Rep} and evaluates to the C identifier of the corresponding run-time array.

\section{The VBR Format}
\label{sec:vbr_format}

This section describes the in-memory representation of the Variable Block Row (VBR) format~\cite{SPARSKIT} targeted by the \lstinline{BlockExtractor} of \autoref{sec:block_partitioner}.
We briefly describe the format in \autoref{fig:vbr}.
When traversing in a row-major order, observe that the following blocks are dense---\colorbox{blue!30}{0}, \colorbox{yellow}{2}, \colorbox{green}{4}, \colorbox{lime}{6}, \colorbox{pink}{7}, \colorbox{cyan}{10}, \colorbox{Salmon}{11}, \colorbox{Orchid}{12}, \colorbox{Apricot}{13}, \colorbox{CadetBlue}{17}, \colorbox{Thistle}{18}, \colorbox{Dandelion}{20}, and \colorbox{Gray}{24}.
\autoref{fig:vbr_indirection} shows the indirection arrays that describe the VBR structure.

\renewcommand{\arraystretch}{1.1}
\begin{figure*}
    \centering
    \begin{adjustbox}{minipage=0.42\linewidth}
    \begin{subfigure}{0.99\textwidth}
        \centering
        \begin{tabular}{| c c | c c c | c | c c c | c c |}
            \hline
            \cellcolor{blue!30}4 & \cellcolor{blue!30}2 & 0 & 0 & 0 & \cellcolor{yellow}1 & 0 & 0 & 0 & \cellcolor{green}-1 & \cellcolor{green}1 \\%[0.3em]
            \cellcolor{blue!30}1 & \cellcolor{blue!30}5 & 0 & 0 & 0 & \cellcolor{yellow}2 & 0 & 0 & 0 & \cellcolor{green}0 & \cellcolor{green}-1 \\%[0.3em]
            \hline
            0 & 0 & \cellcolor{lime}6 & \cellcolor{lime}1 & \cellcolor{lime}2 & \cellcolor{pink}2 & 0 & 0 & 0 & 0 & 0 \\%[0.3em]
            0 & 0 & \cellcolor{lime}2 & \cellcolor{lime}7 & \cellcolor{lime}1 & \cellcolor{pink}0 & 0 & 0 & 0 & 0 & 0 \\%[0.3em]
            0 & 0 & \cellcolor{lime}-1 & \cellcolor{lime}2 & \cellcolor{lime}9 & \cellcolor{pink}3 & 0 & 0 & 0 & 0 & 0 \\%[0.3em]
            \hline
            \cellcolor{cyan}2 & \cellcolor{cyan}1 & \cellcolor{Salmon}3 & \cellcolor{Salmon}4 & \cellcolor{Salmon}5 & \cellcolor{Orchid}10 & \cellcolor{Apricot}4 & \cellcolor{Apricot}3 & \cellcolor{Apricot}2 & 0 & 0 \\%[0.3em]
            \hline
            0 & 0 & 0 & 0 & 0 & \cellcolor{CadetBlue}4 & \cellcolor{Thistle}13 & \cellcolor{Thistle}4 & \cellcolor{Thistle}2 & 0 & 0 \\%[0.3em]
            0 & 0 & 0 & 0 & 0 & \cellcolor{CadetBlue}3 & \cellcolor{Thistle}3 & \cellcolor{Thistle}11 & \cellcolor{Thistle}3 & 0 & 0 \\%[0.3em]
            0 & 0 & 0 & 0 & 0 & \cellcolor{CadetBlue}0 & \cellcolor{Thistle}2 & \cellcolor{Thistle}0 & \cellcolor{Thistle}7 & 0 & 0 \\%[0.3em]
            \hline
            \cellcolor{Dandelion}8 & \cellcolor{Dandelion}4 & 0 & 0 & 0 & 0 & 0 & 0 & 0 & \cellcolor{Gray}25 & \cellcolor{Gray}3 \\%[0.3em]
            \cellcolor{Dandelion}-2 & \cellcolor{Dandelion}3 & 0 & 0 & 0 & 0 & 0 & 0 & 0 & \cellcolor{Gray}8 & \cellcolor{Gray}12 \\%[0.3em]
            \hline
        \end{tabular}
        \caption{Matrix partitioned into Variable block rows}
    \end{subfigure}
    \end{adjustbox}
    \begin{adjustbox}{minipage=0.57\linewidth}
    \begin{subfigure}{0.99\textwidth}
        {\small
        \begin{itemize}
            \item Val: [\colorbox{blue!30}{4, 1, 2, 5}, \colorbox{yellow}{1, 2}, \colorbox{green}{-1, 0, 1, -1}, \colorbox{lime}{6, 2, -1, 1, 7, 2, 2, 1, 9}, \colorbox{pink}{2, 0, 3}, \colorbox{cyan}{2, 1}, \colorbox{Salmon}{3, 4, 5}, \colorbox{Orchid}{10}, \colorbox{Apricot}{4, 3, 2}, \colorbox{CadetBlue}{4, 3, 0}, \colorbox{Thistle}{13, 3, 2, 4, 11, 0, 2, 3, 7}, \colorbox{Dandelion}{8, -2, 4, 3}, \colorbox{Gray}{25, 8, 3, 12}]. Contains the values of the dense blocks in column-major order.
            \item Indx: [\colorbox{blue!30}{0}, \colorbox{yellow}{4}, \colorbox{green}{6}, \colorbox{lime}{10}, \colorbox{pink}{19}, \colorbox{cyan}{22}, \colorbox{Salmon}{24}, \colorbox{Orchid}{27}, \colorbox{Apricot}{28}, \colorbox{CadetBlue}{31}, \colorbox{Thistle}{34}, \colorbox{Dandelion}{43}, \colorbox{Gray}{47}, 51]. Contains the starting index of each dense block in the \lstinline{Val} array. The array is inclusive of the length of the \lstinline{Val} array.
            \item Bindx: [\colorbox{blue!30}{0}, \colorbox{yellow}{2}, \colorbox{green}{4}, \colorbox{lime}{1}, \colorbox{pink}{2}, \colorbox{cyan}{0}, \colorbox{Salmon}{1}, \colorbox{Orchid}{2}, \colorbox{Apricot}{3}, \colorbox{CadetBlue}{2}, \colorbox{Thistle}{3}, \colorbox{Dandelion}{0}, \colorbox{Gray}{4}]. Contains the indices of the blocks in each \textit{block} row.
            \item Rpntr: [0, 2, 5, 6, 9, 11]. Contains the starting index of each block row.
            \item Cpntr: [0, 2, 5, 6, 9, 11]. Contains the starting index of each block column.
            \item Bpntrb: [0, 3, 5, 9, 11, 13]. A CSR-style row pointer into \lstinline{Bindx}: the blocks of block row~$x$ are \lstinline{Bindx[Bpntrb[x]:Bpntrb[x+1]]}, so the difference between consecutive entries gives the number of dense blocks in each block row. The array has length one greater than the number of block rows, and its final entry is the total number of dense blocks.
        \end{itemize}
        }
        \caption{Indirection arrays for the VBR matrix}
        \label{fig:vbr_indirection}
    \end{subfigure}
    \end{adjustbox}
    \caption{Matrix represented in the Variable Block Row format}
    \label{fig:vbr}
\end{figure*}

\section{Block Coverage Threshold}
\label{sec:threshold}

To validate the design of our partitioner, we consider several
variants that change the minimum block-coverage threshold---the
fraction of a matrix's non-zeros that must reside in identified
\ddense{} blocks for \sys{} to accept the matrix for format conversion
(\autoref{sec:block_partitioner}). We ran \sys{} on the full set of 117
SuiteSparse matrices---including the 55 matrices in
\autoref{table:matrices} and 62 additional matrices whose block
coverage fell below 8\%---and varied the acceptance threshold from
0\% to 20\%. \autoref{table:threshold} reports, for each threshold,
the number of included matrices, how many \sys{} outperforms or
underperforms the best fully-sparse baseline, and the geometric-mean
speedup.

With no threshold (0\%), all 117 matrices are included, but roughly a
third exhibit slowdowns (33 for SpMV, 35 for SpMM), dragging the
geometric-mean speedups down to $1.05{\times}$ and $1.11{\times}$. As
the threshold increases, the number of slowdowns drops sharply while
the geometric mean rises. The
8\% threshold strikes a good balance: 52 of 55 matrices show speedup
for SpMV ($1.10{\times}$ geometric mean) and 44 of 55 for SpMM
($1.20{\times}$), while adding only 1 SpMV slowdown
compared to stricter thresholds such as 9\%.
Beyond 12\%, the threshold begins to
exclude matrices that genuinely benefit from \sys{}---at 13\%, the
included set drops sharply from 48 to 33, eliminating the entire
\textit{lowThrust} family (block coverage ${\sim}13\%$) that
consistently shows meaningful speedups (\autoref{table:matrices}).
The 8\% threshold thus
provides a principled cutoff: it filters out matrices whose \ddense{}
blocks cover too few non-zeros to offset the overhead of dense
execution, while retaining nearly all matrices where the approach
yields consistent improvements.

\begin{table}[t]
    \caption{Effect of the minimum block-coverage threshold on \sys{} performance.
    For each threshold (minimum percentage of non-zeros residing in \ddense{} blocks),
    we report the number of included matrices, how many \sys{} outperforms (${>}1{\times}$)
    or underperforms (${\leq}1{\times}$) the best fully-sparse baseline, and the geometric-mean speedup.}
    \label{table:threshold}
    \centering
    \small
    \resizebox{\columnwidth}{!}{
    \begin{tabular}{r|r|rrr|rrr}
    \toprule
    & & \multicolumn{3}{c|}{\textbf{SpMV}} & \multicolumn{3}{c}{\textbf{SpMM}} \\
    \cmidrule(lr){3-5} \cmidrule(l){6-8}
    \textbf{Thr.} & \textbf{\#Mat.}
    & \textbf{${>}1{\times}$} & \textbf{${\leq}1{\times}$} & \textbf{Geo.\ mean}
    & \textbf{${>}1{\times}$} & \textbf{${\leq}1{\times}$} & \textbf{Geo.\ mean} \\
    \midrule
    0\% & 117 & 84 & 33 & 1.05$\times$ & 82 & 35 & 1.11$\times$ \\
    1\% & 94 & 75 & 19 & 1.06$\times$ & 71 & 23 & 1.14$\times$ \\
    2\% & 83 & 71 & 12 & 1.07$\times$ & 66 & 17 & 1.15$\times$ \\
    3\% & 77 & 66 & 11 & 1.07$\times$ & 61 & 16 & 1.16$\times$ \\
    4\% & 70 & 62 & 8 & 1.08$\times$ & 55 & 15 & 1.17$\times$ \\
    5\% & 67 & 60 & 7 & 1.08$\times$ & 52 & 15 & 1.18$\times$ \\
    6\% & 63 & 58 & 5 & 1.09$\times$ & 48 & 15 & 1.18$\times$ \\
    7\% & 60 & 56 & 4 & 1.09$\times$ & 46 & 14 & 1.18$\times$ \\
    \rowcolor{gray!20} 8\% & 55 & 52 & 3 & 1.10$\times$ & 44 & 11 & 1.20$\times$ \\
    9\% & 54 & 51 & 3 & 1.10$\times$ & 44 & 10 & 1.21$\times$ \\
    10\% & 51 & 49 & 2 & 1.11$\times$ & 43 & 8 & 1.22$\times$ \\
    11\% & 51 & 49 & 2 & 1.11$\times$ & 43 & 8 & 1.22$\times$ \\
    12\% & 48 & 46 & 2 & 1.11$\times$ & 41 & 7 & 1.23$\times$ \\
    13\% & 33 & 31 & 2 & 1.15$\times$ & 29 & 4 & 1.30$\times$ \\
    14\% & 31 & 29 & 2 & 1.16$\times$ & 27 & 4 & 1.32$\times$ \\
    15\% & 30 & 29 & 1 & 1.16$\times$ & 26 & 4 & 1.32$\times$ \\
    20\% & 28 & 27 & 1 & 1.17$\times$ & 24 & 4 & 1.30$\times$ \\
    \bottomrule
    \end{tabular}
    }
\end{table}

\section{Band Extractor Algorithms}
\label{sec:bandextractor_algorithms}

This section gives the full algorithms for the \lstinline{BandExtractor}
described in \autoref{sec:band_partitioner}.
Algorithm~\ref{alg:bandextractor} shows the top-level extraction
procedure, and Algorithm~\ref{alg:growsegments} shows the segment-growth
subroutine it invokes.

\begin{algorithm}[t]
\caption{Band Extractor}\label{alg:bandextractor}
\begin{algorithmic}[1]
\Require CSR matrix $A$
\Statex
\Function{ExtractBands}{$A$}
  \State $M[d] \gets$ nnz on diagonal $d$, for $|d| \leq d_{\max}$
  \State $D \gets$ offsets with $M[d] \geq$ count threshold, desc.\ by $M[d]$
  \State $\mathit{used}[0 \ldots n{-}1] \gets \textbf{false}$
  \For{each candidate diagonal $d \in D$}
    \For{each row $i$ with $\lnot\,\mathit{used}[i]$}
      \State $(l_i, u_i, k_i) \gets \Call{FitRibbon}{A, i, d}$
    \EndFor
    \State $S \gets \Call{GrowSegments}{\{(l_i, u_i, k_i)\}, A, d}$
    \If{$S = \emptyset$} \textbf{continue} \EndIf
    \If{$\sum_s \mathrm{area}(s) < A_{\min}$ \textbf{or} $\mathrm{density}(S) < \delta$}
      \State \textbf{continue}
    \EndIf
    \State Accept; $\mathit{used}[r] \gets \textbf{true}$ for all rows in $S$
  \EndFor
  \State \Return accepted regions
\EndFunction
\end{algorithmic}
\end{algorithm}

\begin{algorithm}[t]
\caption{Segment Growth}\label{alg:growsegments}
\begin{algorithmic}[1]
\Require Per-row extents $\{(l_i, u_i, k_i)\}$; CSR matrix $A$; diagonal $d$
\Statex
\Function{GrowSegments}{$\{(l_i, u_i, k_i)\}, A, d$}
  \State $S \gets \emptyset$;\; $i \gets 0$
  \While{$i < n$}
    \If{$k_i = 0$} $i \gets i + 1$; \textbf{continue} \EndIf
    \State $i_0 \gets i$;\; $L \gets l_i$;\; $U \gets u_i$
    \State $z \gets$ nnz in ribbon $[d{-}L,\; d{+}U]$ at row $i_0$;\; $j \gets i{+}1$
    \While{$j < n$ \textbf{and} $k_j > 0$}
      \State $L' \gets \max(L, l_j)$;\; $U' \gets \max(U, u_j)$
      \If{$L' \neq L$ \textbf{or} $U' \neq U$}
        \State $z' \gets$ recount nnz over $[i_0, j]$ with $(L', U')$
      \Else
        \State $z' \gets z +{}$ nnz at row $j$
      \EndIf
      \If{$z' / ((L'{+}U'{+}1)(j{-}i_0{+}1)) < \delta$} \textbf{break} \EndIf
      \State $L \gets L'$;\; $U \gets U'$;\; $z \gets z'$;\; $j \gets j{+}1$
    \EndWhile
    \If{$j - i_0 \geq \ell_{\mathrm{seg}}$} $S \gets S \cup \{(i_0, j, L, U, z)\}$ \EndIf
    \State $i \gets j$
  \EndWhile
  \State \Return $S$
\EndFunction
\end{algorithmic}
\end{algorithm}

\section{Detailed Single-Threaded Results}
\label{sec:detailed_results}

This section expands the single-threaded results of \autoref{sec:rq1}
with the full per-matrix data behind \autoref{fig:bench_executor},
\autoref{fig:vdia_csr_best}, and \autoref{fig:vdia_vbr_csr_best}.
For each of the three format configurations we report every matrix's
salient features together with the running time and speedup of \sys{}
against each individual baseline, for both SpMV and SpMM.
Unless attributed to a specific baseline, the speedups quoted in the prose below are over the best baseline for each matrix, as in the figures of \autoref{sec:rq1}.

\subsection{VBR+CSR}
\label{sec:vbr_csr_detailed}

\begin{table*}[t!]
  \caption{Comparison of \sys{} using VBR+CSR against other approaches for sparse
    matrix computation on 55 SuiteSparse matrices identified by our block partitioner
    as potentially benefiting from our approach. The first group of columns gives the salient
    features of each benchmark: rows,
    columns, non-zeros, number of \ddense{} blocks (\textit{Bl.}),
    average block density (\textit{Avg.\ dens.}), and the percentage of non-zeros
    residing in \ddense{} blocks (\textit{\% Nnz}). The next two groups of
    columns reports the performance improvement of \sys{}  over
    each baseline on SpMV and SpMM computations, respectively (time in $\mu$s and speedup).
    The highest speedup for each benchmark is given in bold; the lowest time is underlined.}
    \label{table:matrices}
    \centering
    \resizebox{\textwidth}{!}{
    \begin{tabular}{l|rrrrrr|rrrr|rrr}
    \toprule
    & & & & & & & \multicolumn{7}{c}{\textbf{\sys speedup over individual baselines}} \\
    \cmidrule(l){8-14}
    & & & & & & & \multicolumn{4}{c|}{\textbf{SpMV}} & \multicolumn{3}{c}{\textbf{SpMM}} \\
    \cmidrule(lr){8-11} \cmidrule(l){12-14}
    \textbf{Matrix} & \textbf{Rows} & \textbf{Cols} & \textbf{Nnz} & \textbf{Bl.} & \textbf{\shortstack{Avg.\\dens. (\%)}} & \textbf{\shortstack{\% Nnz}} & \textbf{SpV8} & \textbf{MKL} & \textbf{Naive} & \textbf{UZP} & \textbf{SpReg} & \textbf{Naive} & \textbf{MKL} \\
    \midrule
    eris1176 & 1,176 & 1,176 & 18,552 & 3 & 72.4 & 56.2 & \underline{6} (1.04$\times$) & 7 (1.09$\times$) & \underline{6} (1.17$\times$) & 11 (\textbf{1.47}$\times$) & 2,691 (1.00$\times$) & 849 (1.59$\times$) & \underline{822} (\textbf{1.62}$\times$) \\
    c-30 & 5,321 & 5,321 & 65,693 & 2 & 94.6 & 41.9 & \underline{18} (1.15$\times$) & 24 (1.09$\times$) & 24 (1.10$\times$) & 42 (\textbf{1.44}$\times$) & 25,903 (1.05$\times$) & \underline{5,058} (1.39$\times$) & 5,486 (\textbf{2.02}$\times$) \\
    lowThrust\_3 & 7,064 & 7,064 & 80,645 & 2 & 100.0 & 13.6 & \underline{26} (1.05$\times$) & 32 (0.99$\times$) & 29 (1.12$\times$) & 49 (\textbf{1.19}$\times$) & 39,524 (1.02$\times$) & 10,512 (0.95$\times$) & \underline{8,708} (\textbf{1.07}$\times$) \\
    jendrec1 & 2,109 & 4,228 & 89,608 & 1 & 55.9 & 88.1 & \underline{23} (1.64$\times$) & 27 (0.96$\times$) & 24 (1.14$\times$) & 27 (\textbf{1.79}$\times$) & 14,105 (0.91$\times$) & 3,065 (\textbf{2.14}$\times$) & \underline{2,926} (2.09$\times$) \\
    orani678 & 2,529 & 2,529 & 90,158 & 1 & 86.1 & 43.4 & \underline{24} (1.21$\times$) & 26 (1.12$\times$) & 26 (1.09$\times$) & 41 (\textbf{1.31}$\times$) & 10,438 (1.02$\times$) & \underline{4,951} (\textbf{1.41}$\times$) & 6,604 (1.32$\times$) \\
    vsp\_c-30\_data\_data & 11,023 & 11,023 & 124,368 & 2 & 94.6 & 22.1 & \underline{46} (1.13$\times$) & 49 (1.02$\times$) & 61 (1.05$\times$) & 94 (\textbf{1.21}$\times$) & 141,436 (1.02$\times$) & 12,722 (1.13$\times$) & \underline{10,363} (\textbf{1.47}$\times$) \\
    lp\_fit2d & 25 & 10,524 & 129,042 & 1 & 100.0 & 8.1 & 43 (\textbf{1.05}$\times$) & 40 (1.03$\times$) & \underline{35} (1.02$\times$) & 177 (1.05$\times$) & \underline{12,827} (0.69$\times$) & 17,919 (1.01$\times$) & 43,260 (\textbf{1.05}$\times$) \\
    lp\_osa\_07 & 1,118 & 25,067 & 144,812 & 1 & 73.5 & 12.2 & 62 (1.01$\times$) & 49 (\textbf{1.01}$\times$) & \underline{46} (1.01$\times$) & 213 (0.96$\times$) & 52,359 (0.92$\times$) & \underline{37,442} (0.99$\times$) & 88,500 (\textbf{1.12}$\times$) \\
    cari & 400 & 1,200 & 152,800 & 1 & 95.0 & 99.5 & 21 (2.40$\times$) & 22 (1.89$\times$) & 20 (2.05$\times$) & \underline{17} (\textbf{4.30}$\times$) & 4,505 (1.51$\times$) & \underline{1,843} (5.95$\times$) & 1,912 (\textbf{6.62}$\times$) \\
    lowThrust\_4 & 13,562 & 13,562 & 160,947 & 2 & 100.0 & 13.1 & 65 (1.04$\times$) & \underline{63} (1.06$\times$) & 66 (1.01$\times$) & 97 (\textbf{1.21}$\times$) & 33,796 (0.97$\times$) & 24,537 (0.95$\times$) & \underline{19,960} (\textbf{1.05}$\times$) \\
    c-45 & 13,206 & 13,206 & 174,452 & 5 & 86.8 & 25.8 & \underline{65} (1.11$\times$) & 69 (1.02$\times$) & \underline{65} (1.02$\times$) & 120 (\textbf{1.28}$\times$) & 206,740 (1.00$\times$) & \underline{19,763} (1.13$\times$) & 23,155 (\textbf{1.36}$\times$) \\
    brainpc2 & 27,607 & 27,607 & 179,395 & 2 & 100.0 & 30.8 & 77 (1.11$\times$) & 97 (0.94$\times$) & \underline{76} (1.06$\times$) & 118 (\textbf{1.60}$\times$) & 823,455 (1.01$\times$) & \underline{36,090} (1.02$\times$) & 37,412 (\textbf{1.26}$\times$) \\
    lowThrust\_5 & 16,262 & 16,262 & 198,369 & 2 & 100.0 & 12.8 & 83 (1.03$\times$) & \underline{77} (1.03$\times$) & 82 (1.01$\times$) & 124 (\textbf{1.20}$\times$) & 304,740 (1.00$\times$) & 31,485 (0.95$\times$) & \underline{26,461} (\textbf{1.05}$\times$) \\
    TSOPF\_RS\_b162\_c1 & 5,374 & 5,374 & 205,399 & 1 & 100.0 & 30.4 & 67 (1.13$\times$) & \underline{61} (1.13$\times$) & \underline{61} (1.10$\times$) & 94 (\textbf{1.24}$\times$) & 76,691 (1.16$\times$) & 12,181 (1.28$\times$) & \underline{11,795} (\textbf{1.29}$\times$) \\
    lowThrust\_6 & 16,928 & 16,928 & 207,349 & 2 & 100.0 & 12.7 & 88 (1.03$\times$) & \underline{81} (1.03$\times$) & 87 (1.00$\times$) & 126 (\textbf{1.20}$\times$) & 336,694 (1.00$\times$) & 33,009 (0.95$\times$) & \underline{27,873} (\textbf{1.05}$\times$) \\
    lowThrust\_7 & 17,378 & 17,378 & 211,561 & 2 & 100.0 & 12.8 & 90 (1.03$\times$) & \underline{83} (1.03$\times$) & 88 (1.00$\times$) & 129 (\textbf{1.22}$\times$) & 192,196 (0.99$\times$) & 34,288 (0.95$\times$) & \underline{29,022} (\textbf{1.04}$\times$) \\
    lowThrust\_8 & 17,702 & 17,702 & 216,445 & 2 & 100.0 & 12.7 & 92 (1.03$\times$) & \underline{85} (1.03$\times$) & 91 (0.99$\times$) & 135 (\textbf{1.20}$\times$) & 52,778 (0.97$\times$) & 34,946 (0.95$\times$) & \underline{29,693} (\textbf{1.04}$\times$) \\
    lowThrust\_9 & 18,044 & 18,044 & 219,589 & 2 & 100.0 & 12.8 & 93 (1.03$\times$) & \underline{86} (1.03$\times$) & 93 (0.97$\times$) & 135 (\textbf{1.21}$\times$) & 379,661 (1.00$\times$) & 36,467 (0.93$\times$) & \underline{30,575} (\textbf{1.04}$\times$) \\
    lowThrust\_10 & 18,260 & 18,260 & 222,005 & 2 & 100.0 & 12.8 & 95 (1.03$\times$) & \underline{87} (1.03$\times$) & 91 (1.02$\times$) & 136 (\textbf{1.21}$\times$) & 81,344 (\textbf{1.04}$\times$) & 36,571 (0.95$\times$) & \underline{30,897} (\textbf{1.04}$\times$) \\
    lowThrust\_11 & 18,368 & 18,368 & 223,801 & 2 & 100.0 & 12.8 & 96 (1.02$\times$) & \underline{88} (1.03$\times$) & 94 (0.99$\times$) & 140 (\textbf{1.20}$\times$) & 60,631 (\textbf{1.03}$\times$) & 36,797 (0.95$\times$) & \underline{31,466} (\textbf{1.03}$\times$) \\
    lowThrust\_12 & 18,458 & 18,458 & 224,593 & 2 & 100.0 & 12.8 & 96 (1.03$\times$) & \underline{88} (1.03$\times$) & 94 (1.01$\times$) & 139 (\textbf{1.20}$\times$) & 199,906 (1.00$\times$) & 37,009 (0.95$\times$) & \underline{30,715} (\textbf{1.03}$\times$) \\
    lowThrust\_13 & 18,476 & 18,476 & 224,897 & 2 & 100.0 & 12.8 & 96 (1.02$\times$) & \underline{89} (1.03$\times$) & 95 (0.99$\times$) & 139 (\textbf{1.20}$\times$) & 398,446 (1.00$\times$) & 37,296 (0.95$\times$) & \underline{30,657} (\textbf{1.06}$\times$) \\
    TSOPF\_RS\_b39\_c7 & 14,098 & 14,098 & 252,446 & 1 & 100.0 & 27.7 & 91 (\textbf{2.21}$\times$) & 87 (1.12$\times$) & \underline{85} (1.07$\times$) & 107 (1.22$\times$) & 240,544 (\textbf{1.00}$\times$) & 26,316 (0.96$\times$) & \underline{23,540} (0.93$\times$) \\
    FX\_March2010 & 1,319 & 9,498 & 301,899 & 1 & 71.0 & 11.2 & 212 (\textbf{1.09}$\times$) & --- & \underline{109} (1.01$\times$) & 323 (1.06$\times$) & \underline{29,688} (0.96$\times$) & 39,575 (1.04$\times$) & 81,238 (\textbf{1.11}$\times$) \\
    lp\_osa\_14 & 2,337 & 54,797 & 317,097 & 1 & 73.1 & 12.1 & 224 (\textbf{1.09}$\times$) & 115 (1.01$\times$) & \underline{112} (1.01$\times$) & 486 (0.96$\times$) & 210,634 (0.95$\times$) & \underline{92,705} (0.98$\times$) & 243,491 (\textbf{1.11}$\times$) \\
    c-57 & 37,833 & 37,833 & 405,197 & 2 & 72.1 & 9.9 & 339 (1.06$\times$) & 191 (0.98$\times$) & \underline{187} (0.99$\times$) & 386 (\textbf{1.08}$\times$) & 179,805 (0.95$\times$) & \underline{74,763} (0.94$\times$) & 90,743 (\textbf{1.02}$\times$) \\
    lp\_osa\_30 & 4,350 & 104,374 & 604,488 & 1 & 72.5 & 12.0 & 485 (\textbf{1.09}$\times$) & 239 (1.01$\times$) & \underline{236} (1.00$\times$) & 966 (0.98$\times$) & 654,797 (1.01$\times$) & \underline{186,154} (0.98$\times$) & 503,332 (\textbf{1.11}$\times$) \\
    TSOPF\_FS\_b162\_c1 & 10,798 & 10,798 & 608,540 & 2 & 100.0 & 20.5 & 428 (1.18$\times$) & \underline{203} (1.07$\times$) & \underline{203} (1.06$\times$) & 340 (\textbf{1.18}$\times$) & 162,399 (1.01$\times$) & \underline{45,473} (1.19$\times$) & 54,672 (\textbf{1.41}$\times$) \\
    TSOPF\_RS\_b162\_c3 & 15,374 & 15,374 & 610,299 & 1 & 100.0 & 30.7 & 409 (\textbf{1.31}$\times$) & 202 (1.11$\times$) & \underline{196} (1.11$\times$) & 299 (1.16$\times$) & 213,423 (1.04$\times$) & 47,611 (\textbf{1.06}$\times$) & \underline{44,704} (1.05$\times$) \\
    heart2 & 2,339 & 2,339 & 682,797 & 12 & 89.3 & 33.2 & 436 (\textbf{1.31}$\times$) & 227 (1.05$\times$) & \underline{225} (1.05$\times$) & 391 (1.20$\times$) & \underline{20,507} (0.99$\times$) & 35,484 (1.37$\times$) & 36,406 (\textbf{1.48}$\times$) \\
    heart3 & 2,339 & 2,339 & 682,797 & 15 & 90.6 & 34.5 & 419 (\textbf{1.32}$\times$) & 227 (1.05$\times$) & \underline{225} (1.05$\times$) & 389 (1.21$\times$) & \underline{19,746} (1.04$\times$) & 35,315 (1.38$\times$) & 35,676 (\textbf{1.51}$\times$) \\
    TSOPF\_RS\_b39\_c19 & 38,098 & 38,098 & 684,206 & 1 & 100.0 & 27.8 & 491 (\textbf{1.25}$\times$) & 253 (1.07$\times$) & \underline{242} (1.07$\times$) & 297 (1.16$\times$) & 821,051 (\textbf{1.01}$\times$) & 74,736 (0.95$\times$) & \underline{66,526} (0.94$\times$) \\
    bundle1 & 10,581 & 10,581 & 770,901 & 17 & 92.4 & 12.2 & 598 (\textbf{1.10}$\times$) & \underline{265} (1.01$\times$) & \underline{265} (1.02$\times$) & 471 (0.95$\times$) & 168,558 (0.99$\times$) & \underline{67,397} (\textbf{1.14}$\times$) & 134,949 (1.07$\times$) \\
    TSOPF\_RS\_b162\_c4 & 20,374 & 20,374 & 812,749 & 1 & 100.0 & 30.8 & 548 (\textbf{1.30}$\times$) & 271 (1.11$\times$) & \underline{264} (1.11$\times$) & 395 (1.17$\times$) & 507,684 (0.99$\times$) & 63,646 (1.06$\times$) & \underline{59,693} (\textbf{1.12}$\times$) \\
    TSC\_OPF\_300 & 9,774 & 9,774 & 820,804 & 2 & 100.0 & 21.5 & 586 (1.20$\times$) & 271 (1.08$\times$) & \underline{270} (1.06$\times$) & 388 (\textbf{1.24}$\times$) & 143,538 (1.05$\times$) & 60,183 (1.08$\times$) & \underline{55,814} (\textbf{1.46}$\times$) \\
    case39 & 40,216 & 40,216 & 1,042,160 & 1 & 100.0 & 9.6 & 871 (\textbf{1.08}$\times$) & 399 (1.04$\times$) & \underline{389} (1.03$\times$) & 597 (1.05$\times$) & 161,904 (\textbf{1.14}$\times$) & \underline{131,433} (0.99$\times$) & 213,920 (1.00$\times$) \\
    connectus & 512 & 394,792 & 1,127,525 & 22 & 100.0 & 23.9 & 1,040 (\textbf{1.16}$\times$) & 618 (1.06$\times$) & \underline{608} (1.05$\times$) & 1,770 (1.11$\times$) & 430,726 (\textbf{1.44}$\times$) & \underline{302,612} (1.10$\times$) & 876,737 (1.27$\times$) \\
    std1\_Jac2 & 21,982 & 21,982 & 1,248,731 & 28 & 62.2 & 14.7 & 987 (1.09$\times$) & 509 (0.99$\times$) & \underline{497} (0.97$\times$) & 867 (\textbf{1.19}$\times$) & 598,588 (1.01$\times$) & \underline{97,147} (\textbf{1.34}$\times$) & 160,722 (1.32$\times$) \\
    heart1 & 3,557 & 3,557 & 1,387,773 & 12 & 86.8 & 16.7 & 1,059 (1.10$\times$) & 517 (1.04$\times$) & \underline{504} (1.05$\times$) & 855 (\textbf{1.12}$\times$) & \underline{43,151} (\textbf{1.90}$\times$) & 93,307 (1.19$\times$) & 120,122 (1.37$\times$) \\
    std1\_Jac3 & 21,982 & 21,982 & 1,455,848 & 27 & 87.9 & 11.7 & 1,213 (1.09$\times$) & 595 (1.03$\times$) & \underline{584} (1.00$\times$) & 1,019 (\textbf{1.12}$\times$) & 601,773 (1.01$\times$) & \underline{115,609} (\textbf{1.34}$\times$) & 221,053 (1.32$\times$) \\
    TSOPF\_RS\_b300\_c1 & 14,538 & 14,538 & 1,474,325 & 1 & 100.0 & 32.3 & 981 (\textbf{1.36}$\times$) & 512 (1.23$\times$) & \underline{505} (1.20$\times$) & 605 (1.19$\times$) & 92,404 (0.99$\times$) & 86,489 (1.25$\times$) & \underline{82,310} (\textbf{1.27}$\times$) \\
    Zd\_Jac2 & 22,835 & 22,835 & 1,642,833 & 34 & 83.7 & 12.5 & 1,350 (1.09$\times$) & 687 (1.07$\times$) & \underline{673} (1.07$\times$) & 1,183 (\textbf{1.15}$\times$) & 604,186 (1.03$\times$) & \underline{121,688} (\textbf{1.36}$\times$) & 224,353 (1.31$\times$) \\
    Zd\_Jac6 & 22,835 & 22,835 & 1,711,983 & 34 & 89.0 & 12.7 & 1,394 (1.10$\times$) & \underline{719} (1.09$\times$) & 728 (\textbf{1.16}$\times$) & 1,221 (1.09$\times$) & 605,426 (1.02$\times$) & \underline{129,535} (\textbf{1.36}$\times$) & 250,010 (1.31$\times$) \\
    Zd\_Jac3 & 22,835 & 22,835 & 1,916,152 & 34 & 89.0 & 11.3 & 1,575 (\textbf{1.09}$\times$) & 859 (1.05$\times$) & \underline{830} (1.06$\times$) & 1,432 (1.06$\times$) & 623,166 (1.00$\times$) & \underline{149,954} (\textbf{1.35}$\times$) & 294,195 (1.32$\times$) \\
    TSC\_OPF\_1047 & 8,140 & 8,140 & 2,016,902 & 3 & 100.0 & 24.1 & 1,451 (\textbf{1.22}$\times$) & 829 (1.16$\times$) & \underline{804} (1.15$\times$) & 877 (1.12$\times$) & \underline{106,466} (1.16$\times$) & 151,364 (1.31$\times$) & 232,945 (\textbf{1.67}$\times$) \\
    net100 & 29,920 & 29,920 & 2,033,200 & 170 & 100.0 & 25.4 & 1,495 (\textbf{1.22}$\times$) & \underline{721} (1.17$\times$) & 728 (1.17$\times$) & 1,246 (1.11$\times$) & 242,951 (1.04$\times$) & \underline{178,983} (\textbf{1.25}$\times$) & 260,104 (1.11$\times$) \\
    gupta1 & 31,802 & 31,802 & 2,164,210 & 6 & 55.7 & 29.3 & 1,669 (\textbf{1.17}$\times$) & \underline{990} (0.95$\times$) & 1,054 (0.96$\times$) & 1,803 (1.11$\times$) & 332,509 (1.11$\times$) & \underline{259,605} (0.95$\times$) & 384,403 (\textbf{1.19}$\times$) \\
    exdata\_1 & 6,001 & 6,001 & 2,269,501 & 1 & 100.0 & 99.3 & 523 (\textbf{3.90}$\times$) & \underline{513} (2.04$\times$) & \underline{513} (2.00$\times$) & 586 (1.91$\times$) & 52,086 (1.38$\times$) & 26,707 (9.72$\times$) & \underline{26,642} (\textbf{25.77}$\times$) \\
    net125 & 36,720 & 36,720 & 2,577,200 & 128 & 100.0 & 19.3 & 2,074 (1.15$\times$) & \underline{1,052} (\textbf{1.20}$\times$) & 1,062 (1.16$\times$) & 1,719 (1.05$\times$) & 313,678 (1.06$\times$) & \underline{230,567} (\textbf{1.25}$\times$) & 328,816 (1.11$\times$) \\
    TSOPF\_RS\_b300\_c2 & 28,338 & 28,338 & 2,943,887 & 1 & 100.0 & 32.3 & 2,054 (\textbf{1.36}$\times$) & 1,364 (1.30$\times$) & 1,376 (1.20$\times$) & \underline{1,195} (1.18$\times$) & 668,747 (1.02$\times$) & 200,495 (\textbf{1.09}$\times$) & \underline{192,684} (1.08$\times$) \\
    net150 & 43,520 & 43,520 & 3,121,200 & 62 & 100.0 & 9.3 & 2,738 (1.07$\times$) & 1,548 (\textbf{1.09}$\times$) & \underline{1,544} (1.07$\times$) & 1,967 (1.04$\times$) & --- & \underline{297,966} (\textbf{1.23}$\times$) & 462,449 (1.06$\times$) \\
    TSOPF\_RS\_b678\_c1 & 18,696 & 18,696 & 4,396,289 & 1 & 100.0 & 32.9 & 3,055 (\textbf{1.29}$\times$) & 2,460 (1.20$\times$) & 2,436 (1.19$\times$) & \underline{1,895} (1.21$\times$) & --- & \underline{237,208} (1.60$\times$) & 314,057 (\textbf{1.74}$\times$) \\
    TSOPF\_RS\_b300\_c3 & 42,138 & 42,138 & 4,413,449 & 1 & 100.0 & 32.4 & 3,151 (\textbf{1.31}$\times$) & 2,551 (1.25$\times$) & 2,628 (1.15$\times$) & \underline{1,859} (1.16$\times$) & --- & 301,039 (\textbf{1.10}$\times$) & \underline{288,687} (1.08$\times$) \\
    TSOPF\_RS\_b678\_c2 & 35,696 & 35,696 & 8,781,949 & 1 & 100.0 & 32.9 & 6,515 (1.20$\times$) & 6,032 (1.18$\times$) & 5,954 (1.17$\times$) & \underline{5,326} (\textbf{1.22}$\times$) & --- & 465,995 (1.57$\times$) & \underline{458,819} (\textbf{2.10}$\times$) \\
    TSOPF\_RS\_b2383 & 38,120 & 38,120 & 16,171,169 & 1 & 100.0 & 33.1 & 12,336 (1.19$\times$) & 11,703 (1.19$\times$) & \underline{11,417} (1.19$\times$) & 12,848 (\textbf{1.22}$\times$) & --- & \underline{1,178,225} (1.60$\times$) & 2,555,522 (\textbf{1.96}$\times$) \\
    \bottomrule
    \end{tabular}
    }
\end{table*}

\autoref{table:matrices} reports the full per-matrix results for the VBR+CSR configuration on the 55 matrices of our benchmark suite.
In general, \sys{} using VBR+CSR improves over every baseline on the majority of benchmarks for both SpMV and SpMM.
The largest gains come from matrices with high block coverage: \textit{exdata\_1} (99.3\% of non-zeros in \ddense{} blocks) achieves $3.90\times$ over SpV8 for SpMV and $25.77\times$ over MKL for SpMM, while \textit{jendrec1} (88.1\%) achieves $1.79\times$ over UZP and $2.14\times$ over naive CSR, respectively.

\paragraph{SpMV}
Matrices with high block coverage (\eg \textit{jendrec1} at 88.1\%, \textit{eris1176} at 56.2\%) show substantial gains over the best fully-sparse baseline.
For matrices with lower block coverage (\eg the \textit{lowThrust} family at ${\sim}$13\%), \sys{} still achieves consistent speedups of $1.02$--$1.05\times$ over the best SpMV baseline by accelerating just the dense portion.
The \textit{lp\_osa} matrices show minimal SpMV improvement ($1.01\times$) because their blocks cover only ${\sim}$12\% of non-zeros and the matrices are highly rectangular, limiting the benefit of dense dispatch.
The three SpMV slowdowns are \textit{c-57} ($0.99\times$), \textit{std1\_Jac2} ($0.97\times$), and \textit{gupta1} ($0.96\times$): \textit{std1\_Jac2}'s blocks average only 62\% density, so the dense kernel spends significant computation on zeros, and the savings from shrinking the sparse problem do not compensate.
The \textit{lp\_osa} matrices have similarly low block density (73\%) but their single-row blocks make the SpMV dense cost negligible, yielding marginal gains ($1.01\times$).

\paragraph{SpMM}
The \textit{Zd\_Jac} and \textit{std1\_Jac} families show consistent SpMM speedups of $1.31$--$1.36\times$ despite modest block coverage ($11$--$15\%$), as SpMM benefits more from the cache locality of dense block storage.
The \textit{heart2} and \textit{heart3} matrices stay within $0.99$--$1.04\times$ of SpReg because their high ratio of non-zeros to matrix columns allows SpReg's packing overhead to be amortized effectively.
The 11 SpMM slowdowns include the \textit{lp\_osa} matrices (${\approx}12\%$ block coverage, ${\approx}73\%$ block density), \textit{lp\_fit2d} ($8.1\%$ coverage, $0.69\times$), \textit{c-57} ($9.9\%$), \textit{case39} ($9.6\%$), and several \textit{TSOPF} matrices whose blocks, despite high density, do not offset the overhead at their coverage levels.
The largest slowdown, \textit{lp\_fit2d}, is a case where SpReg is by far the fastest fully-sparse baseline ($2\times$ faster than naive CSR) but its runtime does not shrink proportionally as non-zeros are extracted: the residual SpReg dispatch runs $13\%$ slower than SpReg on the full matrix despite covering $8\%$ fewer non-zeros, while the dense dispatch adds a further $22\%$ of total time.
\textit{FX\_March2010} ($0.96\times$) slows down the same way: extracting $11\%$ of non-zeros accelerates its SpReg residual by only $6\%$, too little to recover the cost of the dense dispatch.

\subsection{VDIA+CSR}
\label{sec:vdia_csr_detailed}

\begin{table*}[t!]
  \caption{Speedup of VDIA+CSR over CSR on 28 SuiteSparse matrices with diagonal band structure. \textit{Band cov.}\ is the percentage of non-zeros residing in diagonal band segments. \textit{Segs}\ is the number of VDIA segments. Times are in $\mu$s; the highest speedup per benchmark is in bold and the lowest time is underlined; --- marks dispatches not measured for a matrix.}
    \label{table:vdia_csr}
    \centering
    \resizebox{\textwidth}{!}{
    \begin{tabular}{l|rrrrr|rrr|rr}
    \toprule
    & & & & & & \multicolumn{5}{c}{\textbf{VDIA+CSR speedup over CSR}} \\
    \cmidrule(l){7-11}
    & & & & & & \multicolumn{3}{c|}{\textbf{SpMV}} & \multicolumn{2}{c}{\textbf{SpMM}} \\
    \cmidrule(lr){7-9} \cmidrule(l){10-11}
    \textbf{Matrix} & \textbf{Rows} & \textbf{Cols} & \textbf{Nnz} & \textbf{\shortstack{Band\\cov.\ (\%)}} & \textbf{Segs} & \textbf{Naive} & \textbf{MKL} & \textbf{SpV8} & \textbf{Naive} & \textbf{MKL} \\
    \midrule
    vsp\_c-30\_data\_data & 11,023 & 11,023 & 124,368 & 3.9 & 14 & 86 (0.74$\times$) & 67 (0.75$\times$) & \underline{41} (\textbf{1.27}$\times$) & \underline{12,989} (\textbf{1.11}$\times$) & 15,578 (0.97$\times$) \\
    bcsstk28 & 4,410 & 4,410 & 219,024 & 13.6 & 47 & 79 (0.89$\times$) & \underline{75} (\textbf{0.94}$\times$) & 111 (0.76$\times$) & 13,469 (\textbf{1.17}$\times$) & 14,421 (1.05$\times$) \\
    cegb2802 & 2,802 & 2,802 & 277,362 & 3.7 & 5 & \underline{99} (0.93$\times$) & 107 (0.88$\times$) & 101 (\textbf{2.06}$\times$) & 17,598 (\textbf{1.13}$\times$) & 17,765 (1.05$\times$) \\
    cegb2919 & 2,919 & 2,919 & 321,543 & 2.3 & 4 & \underline{119} (0.92$\times$) & 123 (0.90$\times$) & 120 (\textbf{2.02}$\times$) & 20,755 (\textbf{1.15}$\times$) & 21,112 (1.05$\times$) \\
    heart2 & 2,339 & 2,339 & 682,797 & 31.4 & 47 & 223 (1.06$\times$) & 229 (1.04$\times$) & \underline{217} (\textbf{2.63}$\times$) & 31,541 (\textbf{1.55}$\times$) & 36,409 (1.48$\times$) \\
    heart3 & 2,339 & 2,339 & 682,797 & 31.4 & 47 & 223 (1.06$\times$) & 224 (1.07$\times$) & \underline{218} (\textbf{2.54}$\times$) & 31,438 (\textbf{1.55}$\times$) & 35,942 (1.50$\times$) \\
    nemeth19 & 9,506 & 9,506 & 818,302 & 84.1 & 155 & 311 (0.93$\times$) & 317 (0.91$\times$) & \underline{309} (\textbf{2.26}$\times$) & \underline{9,517} (\textbf{6.10}$\times$) & 11,060 (4.92$\times$) \\
    nemeth20 & 9,506 & 9,506 & 971,870 & 0.5 & 2 & \underline{329} (1.08$\times$) & 332 (1.04$\times$) & 334 (\textbf{2.50}$\times$) & 64,527 (\textbf{1.06}$\times$) & 65,021 (0.99$\times$) \\
    olafu & 16,146 & 16,146 & 1,015,156 & 4.4 & 83 & 384 (0.99$\times$) & \underline{363} (0.98$\times$) & 373 (\textbf{2.39}$\times$) & 70,885 (\textbf{1.08}$\times$) & \underline{67,637} (1.04$\times$) \\
    nemeth21 & 9,506 & 9,506 & 1,173,746 & 49.2 & 5 & 450 (0.98$\times$) & 447 (0.97$\times$) & \underline{434} (\textbf{2.28}$\times$) & 39,366 (\textbf{2.08}$\times$) & 44,562 (1.74$\times$) \\
    msc10848 & 10,848 & 10,848 & 1,229,778 & 0.2 & 2 & 428 (1.05$\times$) & 430 (1.04$\times$) & \underline{423} (\textbf{2.52}$\times$) & 89,060 (\textbf{1.05}$\times$) & 94,197 (1.02$\times$) \\
    nemeth22 & 9,506 & 9,506 & 1,358,832 & 92.1 & 156 & \underline{436} (1.21$\times$) & 487 (1.07$\times$) & 461 (\textbf{2.55}$\times$) & \underline{8,075} (\textbf{11.73}$\times$) & 9,915 (9.07$\times$) \\
    heart1 & 3,557 & 3,557 & 1,387,773 & 22.7 & 56 & 458 (1.15$\times$) & 461 (1.16$\times$) & \underline{454} (\textbf{2.56}$\times$) & 79,936 (1.39$\times$) & 117,191 (\textbf{1.40}$\times$) \\
    nemeth24 & 9,506 & 9,506 & 1,506,550 & 0.8 & 3 & \underline{508} (1.18$\times$) & 515 (1.17$\times$) & 514 (\textbf{2.53}$\times$) & 114,854 (\textbf{0.91}$\times$) & 194,630 (0.52$\times$) \\
    nemeth23 & 9,506 & 9,506 & 1,506,810 & 0.8 & 3 & 513 (1.22$\times$) & 510 (1.15$\times$) & \underline{506} (\textbf{2.58}$\times$) & 99,288 (\textbf{1.06}$\times$) & 99,988 (1.01$\times$) \\
    nemeth25 & 9,506 & 9,506 & 1,511,758 & 0.8 & 3 & \underline{506} (1.21$\times$) & 517 (1.17$\times$) & 510 (\textbf{2.52}$\times$) & 110,447 (0.95$\times$) & 100,618 (\textbf{1.00}$\times$) \\
    nemeth26 & 9,506 & 9,506 & 1,511,760 & 0.8 & 3 & \underline{516} (1.17$\times$) & 519 (1.17$\times$) & 517 (\textbf{2.51}$\times$) & 100,107 (1.05$\times$) & 101,449 (\textbf{1.11}$\times$) \\
    opt1 & 15,449 & 15,449 & 1,930,655 & 0.1 & 2 & 666 (1.13$\times$) & \underline{655} (1.15$\times$) & 671 (\textbf{2.57}$\times$) & 133,349 (\textbf{1.06}$\times$) & 136,654 (1.02$\times$) \\
    bcsstk32 & 44,609 & 44,609 & 2,014,701 & 10.3 & 299 & 868 (1.05$\times$) & \underline{825} (1.03$\times$) & 897 (\textbf{2.04}$\times$) & 139,511 (\textbf{1.16}$\times$) & \underline{134,720} (1.07$\times$) \\
    TSC\_OPF\_1047 & 8,140 & 8,140 & 2,016,902 & 12.6 & 33 & 684 (1.43$\times$) & \underline{662} (1.38$\times$) & 674 (\textbf{2.66}$\times$) & 155,909 (1.27$\times$) & 272,729 (\textbf{1.43}$\times$) \\
    pkustk07 & 16,860 & 16,860 & 2,418,804 & 0.5 & 11 & 860 (1.27$\times$) & 855 (1.28$\times$) & \underline{845} (\textbf{2.57}$\times$) & \underline{180,419} (\textbf{1.04}$\times$) & 200,625 (1.03$\times$) \\
    pkustk08 & 22,209 & 22,209 & 3,226,671 & 0.5 & 13 & 1,367 (1.21$\times$) & \underline{1,279} (1.35$\times$) & 1,322 (\textbf{2.25}$\times$) & \underline{246,436} (\textbf{1.03}$\times$) & 275,376 (1.02$\times$) \\
    nd3k & 9,000 & 9,000 & 3,279,690 & 12.9 & 174 & 1,214 (1.66$\times$) & 1,333 (1.52$\times$) & \underline{1,147} (\textbf{2.56}$\times$) & 202,276 (1.28$\times$) & 221,748 (\textbf{1.47}$\times$) \\
    gupta3 & 16,783 & 16,783 & 9,323,427 & 0.1 & 27 & \underline{7,139} (1.03$\times$) & 7,144 (\textbf{1.06}$\times$) & 7,981 (1.05$\times$) & \underline{993,853} (1.00$\times$) & 2,014,770 (\textbf{1.04}$\times$) \\
    ohne2 & 181,343 & 181,343 & 11,063,545 & 1.2 & 424 & --- & \underline{5,474} (\textbf{1.81}$\times$) & 5,659 (1.76$\times$) & --- & 660,124 (1.43$\times$) \\
    memchip & 2,707,524 & 2,707,524 & 14,810,202 & 1.0 & 436 & --- & \underline{16,888} (\textbf{1.07}$\times$) & 20,674 (1.03$\times$) & --- & 3,056,735 (1.02$\times$) \\
    circuit5M\_dc & 3,523,317 & 3,523,317 & 19,194,193 & 0.7 & 506 & --- & \underline{21,412} (\textbf{1.18}$\times$) & 26,656 (1.11$\times$) & --- & 3,795,357 (1.08$\times$) \\
    CoupCons3D & 416,800 & 416,800 & 22,322,336 & 12.9 & 3210 & --- & \underline{18,813} (1.05$\times$) & 19,280 (\textbf{1.08}$\times$) & --- & 1,630,111 (1.11$\times$) \\
    \bottomrule
    \end{tabular}
    }
\end{table*}

\autoref{table:vdia_csr} reports the full per-matrix results for the VDIA+CSR configuration on the 28 matrices with diagonal band structure: the 24 identified by our band extractor, and the four drawn from \citet{fukaya2021mhdc}, which were evaluated with the MKL and SpV8 CSR dispatches for SpMV and the MKL dispatch for SpMM.
In general, \sys{} using VDIA+CSR improves over the CSR baselines on the majority of benchmarks for both SpMV and SpMM, with the SpMM gains tracking band coverage closely and the SpMV gains largely independent of it, as discussed in \autoref{sec:vdia_csr_eval}.

\paragraph{SpMM}
The highest-coverage matrices see the largest gains---\textit{nemeth22} ($92.1\%$ of non-zeros in band segments) reaches $11.14\times$ and \textit{nemeth19} ($84.1\%$) reaches $5.72\times$---as the kernel's reuse across the columns of the dense matrix turns its regular access into far better vectorization than scattered CSR.
Coverage then falls off smoothly through \textit{nemeth21} ($49.2\%$, $1.97\times$), the \textit{heart} family ($22$--$31\%$, ${\sim}1.4$--$1.6\times$), \textit{nd3k} ($12.9\%$, $1.28\times$), and \textit{CoupCons3D} ($12.9\%$, $1.11\times$), down to near-neutral results for the matrices whose bands cover under $1\%$ of non-zeros (\eg \textit{gupta3} at $0.1\%$, $1.00\times$).
The one exception to the trend is \textit{ohne2} ($1.43\times$ at just $1.2\%$ coverage).
The one real slowdown is \textit{nemeth24} ($0.88\times$): its band covers just $0.8\%$ of non-zeros.

\paragraph{SpMV}
The largest gains come from \textit{ohne2} ($1.81\times$, $1.2\%$ coverage), \textit{nd3k} ($1.75\times$, $12.9\%$), and \textit{TSC\_OPF\_1047} ($1.38\times$, $12.6\%$), whereas the $92.1\%$-coverage \textit{nemeth22} gains only $1.20\times$.
Because the diagonal kernel's cache benefit is largely independent of coverage (\autoref{sec:vdia_csr_eval}), even the \textit{nemeth23}--\textit{26} family, whose bands cover under $1\%$ of non-zeros, still gains $1.15$--$1.20\times$.
The six SpMV slowdowns---\textit{cegb2919}, \textit{bcsstk28}, \textit{cegb2802}, \textit{nemeth19}, \textit{olafu}, and \textit{nemeth21}, ranging from $0.92$ to just under $1\times$---are matrices where this cache benefit is too small to offset the cost of splitting the computation into a diagonal pass and a residual CSR pass that each accumulate into the output vector.

\subsection{VDIA+VBR+CSR}
\label{sec:vdia_vbr_csr_detailed}

\begin{table*}[t!]
  \caption{Speedup of VDIA+VBR+CSR over VBR+CSR on 24 SuiteSparse matrices with diagonal band structure. \textit{Band cov.}\ is the percentage of non-zeros residing in diagonal band segments. \textit{Segs}\ is the number of VDIA segments. Times are in $\mu$s; the highest speedup per benchmark is in bold and the lowest time is underlined.}
    \label{table:vdia_vbr_csr}
    \centering
    \resizebox{\textwidth}{!}{
    \begin{tabular}{l|rrrrr|rrr|rr}
    \toprule
    & & & & & & \multicolumn{5}{c}{\textbf{VDIA+VBR+CSR speedup over VBR+CSR}} \\
    \cmidrule(l){7-11}
    & & & & & & \multicolumn{3}{c|}{\textbf{SpMV}} & \multicolumn{2}{c}{\textbf{SpMM}} \\
    \cmidrule(lr){7-9} \cmidrule(l){10-11}
    \textbf{Matrix} & \textbf{Rows} & \textbf{Cols} & \textbf{Nnz} & \textbf{\shortstack{Band\\cov.\ (\%)}} & \textbf{Segs} & \textbf{Naive} & \textbf{MKL} & \textbf{SpV8} & \textbf{Naive} & \textbf{MKL} \\
    \midrule
    vsp\_c-30\_data\_data & 11,023 & 11,023 & 124,368 & 3.9 & 14 & 70 (0.88$\times$) & 60 (0.82$\times$) & \underline{38} (\textbf{1.20}$\times$) & 11,308 (\textbf{1.12}$\times$) & \underline{10,845} (0.96$\times$) \\
    bcsstk28 & 4,410 & 4,410 & 219,024 & 13.6 & 47 & 80 (0.88$\times$) & \underline{79} (\textbf{0.90}$\times$) & 103 (0.81$\times$) & 18,348 (0.85$\times$) & \underline{14,332} (\textbf{1.05}$\times$) \\
    cegb2802 & 2,802 & 2,802 & 277,362 & 3.7 & 5 & \underline{94} (0.97$\times$) & 96 (0.96$\times$) & 103 (\textbf{1.98}$\times$) & 18,476 (\textbf{1.04}$\times$) & 19,984 (0.91$\times$) \\
    cegb2919 & 2,919 & 2,919 & 321,543 & 2.3 & 4 & 120 (0.90$\times$) & \underline{117} (0.93$\times$) & 120 (\textbf{1.95}$\times$) & \underline{21,071} (\textbf{1.08}$\times$) & 22,767 (0.95$\times$) \\
    heart2 & 2,339 & 2,339 & 682,797 & 31.4 & 47 & 256 (0.88$\times$) & 263 (0.86$\times$) & \underline{255} (\textbf{1.71}$\times$) & 28,795 (\textbf{1.23}$\times$) & 30,329 (1.20$\times$) \\
    heart3 & 2,339 & 2,339 & 682,797 & 31.4 & 47 & 261 (0.86$\times$) & 261 (0.87$\times$) & \underline{257} (\textbf{1.63}$\times$) & 28,594 (\textbf{1.24}$\times$) & 30,428 (1.17$\times$) \\
    nemeth19 & 9,506 & 9,506 & 818,302 & 84.1 & 155 & 305 (0.95$\times$) & 339 (0.85$\times$) & \underline{304} (\textbf{2.29}$\times$) & \underline{9,798} (\textbf{5.96}$\times$) & 11,202 (4.82$\times$) \\
    nemeth20 & 9,506 & 9,506 & 971,870 & 0.5 & 2 & \underline{333} (1.05$\times$) & 334 (1.03$\times$) & 338 (\textbf{2.47}$\times$) & 65,851 (\textbf{1.04}$\times$) & 68,251 (0.94$\times$) \\
    olafu & 16,146 & 16,146 & 1,015,156 & 4.4 & 83 & 385 (0.99$\times$) & \underline{365} (0.99$\times$) & 395 (\textbf{2.25}$\times$) & \underline{71,102} (\textbf{1.08}$\times$) & 71,677 (0.97$\times$) \\
    nemeth21 & 9,506 & 9,506 & 1,173,746 & 49.2 & 5 & 451 (0.99$\times$) & 447 (0.98$\times$) & \underline{429} (\textbf{2.30}$\times$) & 40,330 (\textbf{2.08}$\times$) & 43,119 (1.80$\times$) \\
    msc10848 & 10,848 & 10,848 & 1,229,778 & 0.2 & 2 & \underline{426} (1.07$\times$) & 430 (1.05$\times$) & 426 (\textbf{2.47}$\times$) & 88,935 (\textbf{1.06}$\times$) & 94,881 (1.01$\times$) \\
    nemeth22 & 9,506 & 9,506 & 1,358,832 & 92.1 & 156 & 437 (1.21$\times$) & 488 (1.07$\times$) & \underline{432} (\textbf{2.72}$\times$) & \underline{8,571} (\textbf{11.10}$\times$) & 10,036 (9.05$\times$) \\
    heart1 & 3,557 & 3,557 & 1,387,773 & 22.7 & 56 & 483 (1.04$\times$) & 493 (1.05$\times$) & \underline{479} (\textbf{2.21}$\times$) & 76,831 (1.21$\times$) & 88,635 (\textbf{1.36}$\times$) \\
    nemeth24 & 9,506 & 9,506 & 1,506,550 & 0.8 & 3 & \underline{505} (1.20$\times$) & 518 (1.15$\times$) & 510 (\textbf{2.54}$\times$) & 100,713 (\textbf{1.05}$\times$) & 105,326 (0.95$\times$) \\
    nemeth23 & 9,506 & 9,506 & 1,506,810 & 0.8 & 3 & \underline{508} (1.24$\times$) & 526 (1.11$\times$) & 509 (\textbf{2.56}$\times$) & 100,319 (\textbf{1.05}$\times$) & 110,510 (0.90$\times$) \\
    nemeth25 & 9,506 & 9,506 & 1,511,758 & 0.8 & 3 & \underline{511} (1.20$\times$) & 519 (1.13$\times$) & 512 (\textbf{2.50}$\times$) & 104,543 (\textbf{1.01}$\times$) & 124,991 (0.81$\times$) \\
    nemeth26 & 9,506 & 9,506 & 1,511,760 & 0.8 & 3 & 511 (1.17$\times$) & 517 (1.17$\times$) & \underline{508} (\textbf{2.54}$\times$) & 102,027 (1.05$\times$) & 105,739 (\textbf{1.08}$\times$) \\
    opt1 & 15,449 & 15,449 & 1,930,655 & 0.1 & 2 & 669 (1.10$\times$) & 677 (1.12$\times$) & \underline{668} (\textbf{2.55}$\times$) & 137,405 (\textbf{1.05}$\times$) & 138,360 (1.00$\times$) \\
    bcsstk32 & 44,609 & 44,609 & 2,014,701 & 10.3 & 299 & \underline{872} (1.03$\times$) & 885 (0.98$\times$) & 875 (\textbf{2.09}$\times$) & 142,730 (\textbf{1.17}$\times$) & \underline{135,532} (1.06$\times$) \\
    TSC\_OPF\_1047 & 8,140 & 8,140 & 2,016,902 & 12.6 & 33 & 631 (1.27$\times$) & \underline{617} (1.26$\times$) & 626 (\textbf{2.34}$\times$) & 109,439 (1.38$\times$) & 127,924 (\textbf{1.82}$\times$) \\
    pkustk07 & 16,860 & 16,860 & 2,418,804 & 0.5 & 11 & \underline{878} (1.21$\times$) & 926 (1.17$\times$) & 891 (\textbf{2.36}$\times$) & \underline{177,200} (0.79$\times$) & 213,718 (\textbf{0.94}$\times$) \\
    pkustk08 & 22,209 & 22,209 & 3,226,671 & 0.5 & 13 & \underline{1,313} (1.28$\times$) & 1,339 (1.26$\times$) & 1,365 (\textbf{2.12}$\times$) & \underline{238,349} (0.80$\times$) & 270,291 (\textbf{1.02}$\times$) \\
    nd3k & 9,000 & 9,000 & 3,279,690 & 12.9 & 174 & 1,470 (1.35$\times$) & \underline{1,335} (1.53$\times$) & 1,532 (\textbf{1.89}$\times$) & 205,067 (0.94$\times$) & 222,116 (\textbf{1.41}$\times$) \\
    gupta3 & 16,783 & 16,783 & 9,323,427 & 0.1 & 27 & \underline{7,189} (1.02$\times$) & 7,515 (1.01$\times$) & 8,006 (\textbf{1.05}$\times$) & \underline{953,664} (\textbf{1.06}$\times$) & 2,042,735 (1.03$\times$) \\
    \bottomrule
    \end{tabular}
    }
\end{table*}

\autoref{table:vdia_vbr_csr} reports the full per-matrix results for the VDIA+VBR+CSR configuration.
The trends mirror those of \autoref{table:vdia_csr}, but against the stronger VBR+CSR baseline, which already dispatches \ddense{} blocks to efficient dense kernels.
The SpMM gains remain concentrated in the matrices with high band coverage: \textit{nemeth22} ($92.1\%$) reaches $10.60\times$ over the best VBR+CSR baseline and \textit{nemeth19} ($84.1\%$) reaches $5.51\times$.

Six matrices---\textit{bcsstk28}, \textit{cegb2919}, \textit{heart2}, \textit{heart3}, \textit{nemeth19}, and \textit{bcsstk32}---slow down for SpMV ($0.88$--$0.99\times$) but speed up for SpMM ($1.03$--$5.51\times$).
These matrices have moderate-to-high band coverage, so the diagonal SpMM kernel still adds value beyond the \ddense{} blocks VBR already captured.
For SpMV the same band cannot help: the VBR+CSR baseline already dispatches the \ddense{} blocks to a fast GEMV kernel, and the overhead of the extra diagonal pass and the residual CSR pass outweighs the band's marginal cache benefit.
Conversely, seven matrices---\textit{vsp\_c-30\_data\_data}, \textit{nemeth20}, \textit{nemeth23}, \textit{nemeth25}, \textit{pkustk07}, \textit{pkustk08}, and \textit{nd3k}---speed up for SpMV but slow down for SpMM.
All but \textit{nd3k} have very low band coverage (${\leq}0.8\%$), and \textit{nd3k}'s band ($12.9\%$) overlaps a matrix whose \ddense{} blocks VBR already extracts.
VBR+CSR's SpMM dispatches these \ddense{} blocks to BLAS GEMM and already runs efficiently, so layering a small diagonal kernel on top only adds dispatch overhead; the SpMM slowdowns are largest where band coverage is lowest (\textit{pkustk07} $0.79\times$, \textit{pkustk08} $0.80\times$).
SpMV still benefits from the diagonal kernel regularizing the residual's access pattern, so it improves even where SpMM does not.
\textit{olafu} and \textit{cegb2802} regress for both kernels ($0.97$--$0.99\times$).
Their bands cover only $4.4\%$ and $3.7\%$ of non-zeros, and VBR+CSR already captures their structure well, so the diagonal extraction contributes too little to either kernel to recover its overhead.

\section{Parallelization Threshold}
\label{sec:par_threshold}

The single-threaded evaluation uses an 8\% block-coverage threshold
(\autoref{sec:threshold}), but parallelization introduces additional
overhead---thread synchronization, cache contention, and load
imbalance---that requires stricter filtering.  We investigate two
complementary filters: raising the minimum block-coverage threshold
and imposing a maximum block aspect ratio.

\paragraph{Coverage threshold alone}
\autoref{table:par_coverage} shows the effect of the block-coverage
threshold on 8-thread parallel performance with no aspect-ratio
filter.
At the single-threaded threshold of 8\%, SpMM performance is strongly
net-negative: only 24 of 55 matrices show speedup (geometric mean
$0.91{\times}$), because many matrices with small block coverage incur
threading overhead that exceeds the parallel BLAS benefit.
Raising the threshold to 15\% flips SpMM to net-positive
($1.04{\times}$ geometric mean, 19 of 30 wins), but 11 matrices still
show slowdowns.  SpMV is less affected because its lower arithmetic
intensity means the sparse dispatch dominates total time regardless.

\paragraph{Aspect ratio as a secondary filter}
Among the 30 matrices passing the 15\% coverage threshold, the
remaining SpMM slowdowns are concentrated in matrices with extremely
elongated blocks (aspect ratio $>100$, \ie
$\max(\mathrm{rows},\mathrm{cols})/\min(\mathrm{rows},\mathrm{cols})$).
\autoref{table:par_aspect} shows the effect of adding an aspect-ratio
cap on top of the 15\% coverage threshold.
Without the aspect filter, 11 of 30 matrices show SpMM slowdowns;
adding a cap of ${\leq}100$ eliminates all but 1 slowdown and raises
the geometric mean from $1.04{\times}$ to $1.48{\times}$.
The 10 excluded matrices---which include configurations such as
\texttt{TSOPF\_RS\_b39\_c19} (aspect $1900$) and \texttt{connectus}
(aspect $75{,}443$)---all have SpMM speedups below $1{\times}$,
confirming that parallel BLAS cannot efficiently exploit such
elongated blocks.
At ${\leq}150$ the same 20 matrices are retained as at ${\leq}100$
(no matrix in our set has worst-case aspect between 100 and 150), so
we adopt ${\leq}100$ as a conservative bound.

\begin{table}[t]
    \caption{Effect of the block-coverage threshold on 8-thread
    parallel performance (no aspect-ratio filter).
    Columns are analogous to \autoref{table:threshold} but report
    speedup of \sys{} over the best fully-sparse baseline at 8 threads.}
    \label{table:par_coverage}
    \centering
    \small
    \resizebox{\columnwidth}{!}{
    \begin{tabular}{r|r|rrr|rrr}
    \toprule
    & & \multicolumn{3}{c|}{\textbf{SpMV}} & \multicolumn{3}{c}{\textbf{SpMM}} \\
    \cmidrule(lr){3-5} \cmidrule(l){6-8}
    \textbf{Thr.} & \textbf{\#Mat.}
    & \textbf{${>}1{\times}$} & \textbf{${\leq}1{\times}$} & \textbf{Geo.\ mean}
    & \textbf{${>}1{\times}$} & \textbf{${\leq}1{\times}$} & \textbf{Geo.\ mean} \\
    \midrule
    0\% & 55 & 34 & 21 & 1.17$\times$ & 24 & 31 & 0.91$\times$ \\
    5\% & 55 & 34 & 21 & 1.17$\times$ & 24 & 31 & 0.91$\times$ \\
    8\% & 55 & 34 & 21 & 1.17$\times$ & 24 & 31 & 0.91$\times$ \\
    10\% & 51 & 33 & 18 & 1.19$\times$ & 24 & 27 & 0.91$\times$ \\
    12\% & 48 & 32 & 16 & 1.20$\times$ & 22 & 26 & 0.90$\times$ \\
    \rowcolor{gray!20} 15\% & 30 & 19 & 11 & 1.23$\times$ & 19 & 11 & 1.04$\times$ \\
    18\% & 29 & 19 & 10 & 1.25$\times$ & 18 & 11 & 1.04$\times$ \\
    20\% & 28 & 19 & 9 & 1.27$\times$ & 18 & 10 & 1.04$\times$ \\
    \bottomrule
    \end{tabular}
    }
\end{table}

\begin{table}[t]
    \caption{Effect of adding a maximum block aspect ratio (max dimension
    $/$ min dimension) on top of the 15\% coverage threshold, at 8 threads.
    The aspect-ratio filter primarily improves SpMM by excluding matrices
    whose elongated blocks cannot be parallelized efficiently.}
    \label{table:par_aspect}
    \centering
    \small
    \resizebox{\columnwidth}{!}{
    \begin{tabular}{r|r|rrr|rrr}
    \toprule
    & & \multicolumn{3}{c|}{\textbf{SpMV}} & \multicolumn{3}{c}{\textbf{SpMM}} \\
    \cmidrule(lr){3-5} \cmidrule(l){6-8}
    \textbf{Max AR} & \textbf{\#Mat.}
    & \textbf{${>}1{\times}$} & \textbf{${\leq}1{\times}$} & \textbf{Geo.\ mean}
    & \textbf{${>}1{\times}$} & \textbf{${\leq}1{\times}$} & \textbf{Geo.\ mean} \\
    \midrule
    none & 30 & 19 & 11 & 1.23$\times$ & 19 & 11 & 1.04$\times$ \\
    ${\leq}200$ & 22 & 14 & 8 & 1.24$\times$ & 19 & 3 & 1.35$\times$ \\
    \rowcolor{gray!20} ${\leq}100$ & 20 & 13 & 7 & 1.26$\times$ & 19 & 1 & 1.48$\times$ \\
    ${\leq}75$ & 15 & 9 & 6 & 1.30$\times$ & 14 & 1 & 1.54$\times$ \\
    ${\leq}50$ & 13 & 7 & 6 & 1.34$\times$ & 12 & 1 & 1.52$\times$ \\
    \bottomrule
    \end{tabular}
    }
\end{table}

\section{Thread Scaling}
\label{sec:thread_scaling}

This section expands the parallel results of \autoref{sec:rq2}, which
report speedups at 8 threads, with a per-matrix analysis of the SpMV
slowdowns at 8 threads and the scaling behavior of \sys{} and the
fully-sparse baselines across thread counts.

\paragraph{SpMV slowdowns at 8 threads}
The seven SpMV slowdowns in \autoref{fig:parallel_spmv} fall into two groups.
The \textit{heart} family (\textit{heart1}--\textit{3}, $0.80$--$0.83\times$) and the two large single-block \textit{TSOPF\_RS\_b678} matrices (both $0.92\times$) have only moderate block coverage ($16$--$35\%$), so the sparse residual dominates total runtime; the fully-sparse baselines parallelize that residual competitively, leaving little room for \sys{}'s parallel GEMV to add value.
\textit{net100} ($0.73\times$) and \textit{net125} ($0.77\times$) instead have many small blocks (170 and 128 blocks, each spanning only ${\sim}33$--$115$ rows), whose per-block GEMV work is too small to amortize the threading overhead of a parallel BLAS call.

\begin{figure*}[t!]
    \centering
    \includegraphics[width=\textwidth]{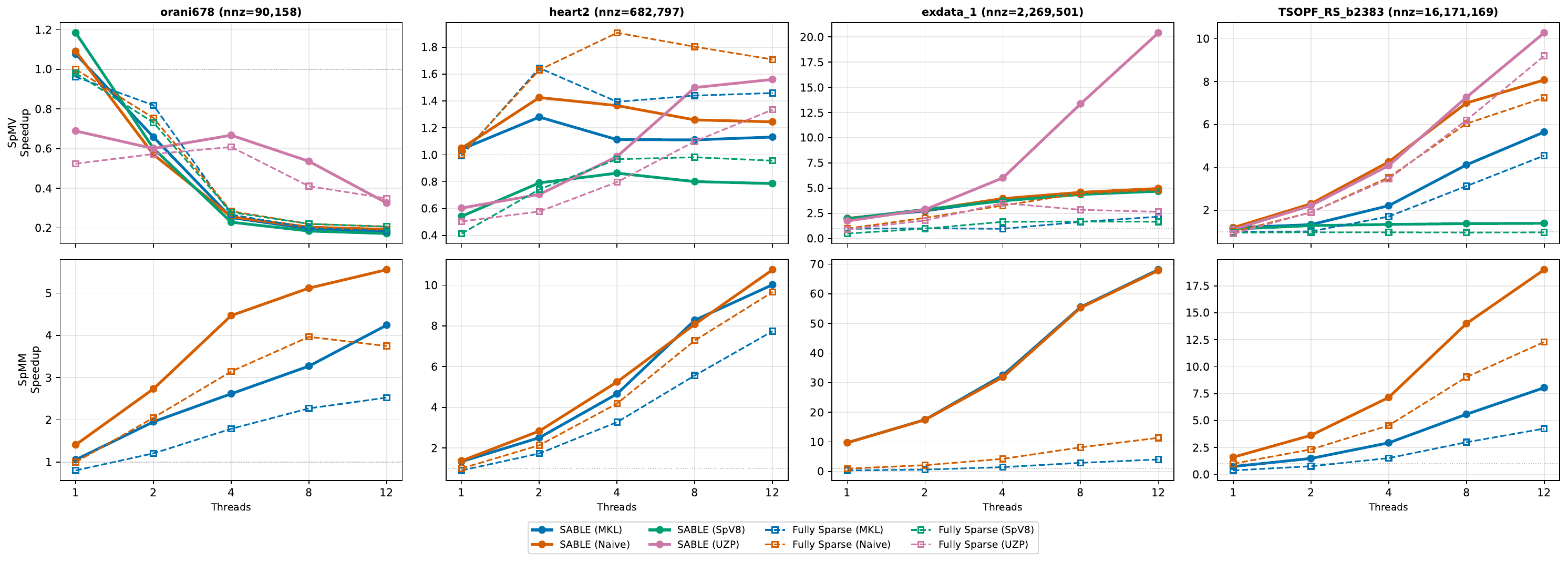}
    \caption{Thread scaling for four representative matrices: SpMV (top row) and SpMM (bottom row).
    Solid lines show \sys{} with different sparse dispatches; dashed lines show the corresponding fully-sparse baselines.
    All values are speedups over the single-threaded time of the best fully-sparse baseline (higher is better).}
    \label{fig:spmv_scaling}
\end{figure*}

\paragraph{Scaling across thread counts}
\autoref{fig:spmv_scaling} shows the thread scaling behavior of all
\sys{} sparse dispatches and fully-sparse baselines for four
representative matrices, with SpMV in the top row and SpMM in the bottom row.
For \textit{orani678} (90K nnz, 43.3\% in blocks), parallelism overhead dominates SpMV due to the small matrix size, and all configurations degrade beyond 2 threads; however, SpMM's higher arithmetic intensity allows \sys{} to maintain an advantage, reaching $5.6\times$ at 12 threads versus $3.7\times$ for fully-sparse.
For \textit{heart2} (683K nnz, 33.3\% in blocks, 12 blocks), the fully-sparse baseline slightly outpaces \sys{} for SpMV at higher thread counts ($1.7\times$ vs $1.6\times$ at 12 threads); for SpMM, however, \sys{} consistently leads, reaching $10.8\times$ versus $9.7\times$ at 12 threads, as the dense blocks provide additional parallelism on top of the sparse dispatch.
\textit{exdata\_1} (2.3M nnz, 99.3\% in blocks) shows the most dramatic scaling: since nearly all computation is dispatched to dense BLAS, \sys{} achieves ${\sim}20\times$ SpMV and ${\sim}68\times$ SpMM speedup at 12 threads, while fully-sparse baselines plateau at $5\times$ and $11\times$, respectively.
For \textit{TSOPF\_RS\_b2383} (16.2M nnz, 33.1\% in blocks), both \sys{} and the fully-sparse baselines scale near-linearly; \sys{} maintains a consistent advantage, reaching $10.3\times$ versus $9.2\times$ for SpMV and $19.0\times$ versus $12.3\times$ for SpMM at 12 threads.

\end{document}